\documentclass[12pt]{article} 
\pdfoutput=1
\usepackage{epsfig,graphicx,psfrag}
\usepackage{amsmath,amsfonts,stmaryrd,latexsym,amssymb}
\usepackage{url,hyperref,cite,relsize}
\newcommand{\be}{\begin{equation}}
\newcommand{\ee}{\end{equation}}
\newcommand{\bea}{\begin{eqnarray}}
\newcommand{\eea}{\end{eqnarray}}
\newcommand{\bear}{\begin{eqnarray}}
\newcommand{\eear}{\end{eqnarray}}

\newcommand{\ba}{\begin{array}}
\newcommand{\ea}{\end{array}}

\newcommand{\bpm}{\begin{pmatrix}}
\newcommand{\epm}{\end{pmatrix}}

\DeclareMathOperator{\Tr}{Tr}

\begin{document}

\baselineskip=18pt \pagestyle{plain} \setcounter{page}{1}

\vspace*{1.5cm}

\begin{center}

{\Large \bf  Top Seesaw with a Custodial Symmetry,\\ and the 126 GeV Higgs} \\ [9mm]

{\normalsize \bf Hsin-Chia Cheng and Jiayin Gu \\ [4mm]
{\small {\it
Department of Physics, University of California, Davis, CA 95616, USA
}}\\
}

\end{center}

\vspace*{0.2cm}

\begin{abstract}
The composite Higgs models based on the top seesaw mechanism commonly possess an enhanced approximate chiral symmetry, which is spontaneously broken to produce the Higgs field as the pseudo-Nambu-Goldstone bosons. The minimal model with only one extra vector-like singlet quark that mixes with the top quark can naturally give rise to a $126$~GeV Higgs boson. However, without having a custodial symmetry it suffers from the weak-isospin violation constraint, which pushes the chiral symmetry breaking scale above a few TeV, causing a substantial fine-tuning for the weak scale. We consider an extension to the minimal model to incorporate the custodial symmetry by adding a vector-like electroweak doublet of quarks with hypercharge +7/6. Such a setup also protects the $Zb\bar{b}$ coupling which is another challenge for many composite Higgs models. With this addition, the chiral symmetry breaking scale can be lowered to around 1 TeV, making the theory much less fine-tuned. The Higgs is a pseudo-Nambu-Goldstone boson of the broken $O(5)$ symmetry. For the Higgs mass to be 126 GeV, the hypercharge $+7/6$ quarks should be around or below the chiral symmetry breaking scale, and are likely to be the lightest new states. The $14$~TeV LHC will significantly extend the search reach of these quarks. To probe the rest of the spectrum, on the other hand, would require a higher-energy future collider.
\end{abstract}


\newpage
{
\tableofcontents}
\newpage

\section{Introduction}
\label{sec:intro}

The nature and properties of the Higgs boson have become the focus of particle physics research since its discovery in 2012. The relatively light Higgs boson of 126 GeV suggests that it is either an elementary particle or a pseudo-Nambu-Goldstone boson (pNGB) of some spontaneously broken symmetry if it is a composite degree of freedom of some strong dynamics~\cite{Kaplan:1983fs, Kaplan:1983sm, Banks:1984gj, Georgi:1984ef, Dugan:1984hq, Koulovassilopoulos:1993pw}. Other than the Higgs boson, the Large Hadron Collider (LHC) so far has not discovered any new physics yet. The couplings of the Higgs boson are consistent with their standard model (SM) values, though some significant deviations are still possible. If there exists new physics responsible for the origin of the electroweak symmetry breaking (EWSB), the current experimental results indicate that it is probably close to the decoupling limit. On the other hand, the naturalness argument strongly prefers the new physics to be near the weak scale to avoid excessive fine-tuning. The tension between these two requirements has becomes a severe challenge for any model that attempts to explain the electroweak (EW) scale.

In a previous paper~\cite{Cheng:2013qwa}, it was found that in a top seesaw model of dynamical EWSB~\cite{Dobrescu:1997nm, Chivukula:1998wd, Dobrescu:1999gv, He:2001fz}, the Higgs boson arises naturally as a pNGB of the spontaneously broken $U(3)_L$ symmetry, which relates the left-handed top-bottom doublet and a new quark $\chi_L$.\footnote{A similar and independent study can be found in Ref.~\cite{Fukano:2013aea}.  The 126 GeV Higgs was also considered  in a top and bottom seesaw model from supersymmetric strong dynamics through tuning and mixing~\cite{Balazs:2012iz}. }  The top seesaw model fixes the problem of the top quark being too heavy in the top condensation model~\cite{Nambu:1989jt, Miransky:1988xi, Miransky:1989ds, Marciano:1989xd, Bardeen:1989ds} by mixing the top quark with a new vector-like quark $\chi$. It was shown that, in the presence of the approximate $U(3)_L$ symmetry, the Higgs boson mass is highly correlated with, and generically smaller than the top quark mass. The experimental value of 126 GeV can be obtained with natural values of the parameters of this model.  A drawback of this model is that the $U(3)_L$ does not contain a custodial symmetry. As a result, the constraint on the weak-isospin violation requires the chiral symmetry breaking scale $f$ to be above 3.5 TeV. Some significant fine-tuning is needed to obtain the weak scale at $v \approx 246$~GeV. Such a high chiral symmetry breaking scale also implies that none of the new states are predicted to be reachable at the LHC. A collider of much higher center of mass energy ($\sim 100$~TeV) would be  needed to have any chance to see some of the new states.
 
It is therefore desirable to consider extensions of the minimal top seesaw model to include a custodial symmetry.
A straightforward extension to the top seesaw model in Ref.~\cite{Cheng:2013qwa} is to introduce ``bottom seesaw" by adding a vector-like singlet bottom partner $\omega$.  The spontaneously broken $U(4)_L$ symmetry can produce 2 light Higgs doublets. Without additional contributions, the mass of the Higgs boson made of the bottom and $\omega$ quarks is related to the bottom quark mass and hence is too light. To avoid this situation, one could introduce scalar mass terms (which come from 4-fermion interactions in the UV theory) to explicitly break the $U(4)_L$ chiral symmetry of $(t_L, b_L, \chi_L, \omega_L)$ down to $Sp(4)$. While the $Sp(4)$ contains the $SU(2)_C$ custodial symmetry which can be used to protect the weak-isospin, such a model suffers from the constraint on $Z\to b\bar{b}$ branching ratio.  The most recent results suggest that the SM prediction for $Z\to b \bar{b}$ branching ratio ($R_b$) is $2.4\sigma$ smaller than the measured value~\cite{Baak:2012kk}.  When the bottom quark mixes with a heavy singlet, as required for the bottom seesaw mechanism, the $Zb_L\bar{b}_L$ coupling is reduced (becomes less negative) while the $Zb_R\bar{b}_R$ coupling is not modified.  As a result, the $Z\to b \bar{b}$ branching ratio is further reduced.   This puts a constraint on the mixing angle ($\theta^b_L$) between $b_L$ and $\omega_L$, which pushes the mass of $\omega$ to be very large \cite{He:2001fz, Collins:1999rz}.  In order not to have a large weak-isospin violation, the masses of $\chi$ and  $\omega$ should be close, again implying a large chiral symmetry breaking scale. By playing with the model parameters, the chiral symmetry breaking scale may only be slightly reduced compared to the original top seesaw model, which means that such an extension still require stiff fine-tunings.

It was pointed out in Ref.~\cite{Agashe:2006at} that the custodial symmetry which protects the weak isospin can also protect the $Zb_L\bar{b}_L$ coupling under certain conditions.  Namely, the new physics needs to be invariant under an $O(4)$ global symmetry, which is the familiar $SU(2)_L\times SU(2)_R$ of the SM Higgs sector together with a parity defined as the interchange $L \leftrightarrow R$ ($P_{LR}$);  also, $b_L$ needs to be charged under both $SU(2)_L$ and $SU(2)_R$ with $T_L=T_R=1/2, \, T^3_L=T^3_R=-1/2\,$.  This implies that the SM $(t_L, b_L)$ doublet needs to be embedded  into a $({\bf 2}, {\bf 2})$ representation of $SU(2)_L\times SU(2)_R$, together with a new doublet quark $(X_L, T_L)$ of hypercharge $+7/6$, with the quantum numbers given in Table~\ref{tab:XTtb}. 
To adopt this setup we introduce an $SU(2)_W$-doublet vector-like quarks, $Q\equiv(X,T)$, with hypercharge $+7/6$, in addition to the vector-like $SU(2)_W$-singlet quark $\chi$ which is responsible for the top seesaw mechanism. 
\begin{table}
\centering
\begin{tabular}{c|cccc}
& $X_L$ & $T_L$ & $t_L$ & $b_L$  \\\hline
$T^3_L$ & 1/2 & -1/2 & 1/2 & -1/2 \\
$T^3_R$ & 1/2 & 1/2 & -1/2 & -1/2 
\end{tabular}
\caption{The quantum numbers of $X_L$, $T_L$, $t_L$, $b_L$ under $SU(2)_L \times SU(2)_R$.}
\label{tab:XTtb}
\end{table}

The underlying strong dynamics is assumed to approximately respect the $U(5)_L\times U(4)_R$ symmetry among the five left-handed quarks $(t_L, b_L, X_L, T_L, \chi_L)$ and the four right-handed quarks $(X_R, T_R, t_R, \chi_R)$. To avoid too many light pNGBs after the chiral symmetry breaking, gauge invariant scalar mass terms (arising from 4-fermion interactions in the UV) can be introduced to explicitly break $U(4)_R$ symmetry and also $U(5)_L$ down to $O(5)$. In this way, only one light Higgs doublet arises from the chiral symmetry breaking of $O(5) \to O(4)$.
An important difference between our model and the setup in Ref.~\cite{Agashe:2006at} is that in our model, the custodial symmetry that protects both the weak isospin and the $Zb_L\bar{b}_L$ coupling is only approximately preserved by the new physics, which violates the conditions in Ref.~\cite{Agashe:2006at}.  Nevertheless, we found that within some regions of the parameter space, both corrections are within experimental constraints, while the chiral symmetry breaking can be as low as $\sim 1$~TeV, significantly ameliorating the fine-tuning of the weak scale.

The rest of this paper is organized as follows.  In Section~\ref{sec:fullmodel}, we write down the effective theory with composite scalars below the compositeness scale with $U(5)_L\times U(4)_R$ symmetric dynamics of the extended quark sector.  In Section~\ref{sec:let}, we focus on the theory at the TeV scale and show that the Higgs boson arises as a pNGB of the chiral symmetry breaking.   We derive an approximate analytic formula for the mass of the Higgs boson ($M_h$) and discuss various possible corrections. It can naturally be around $126$~GeV for model parameters within reasonable ranges.  In Section~\ref{sec:plots}, we further verify the results in Section~\ref{sec:let} with numerical studies.  We show that in this model the chiral symmetry breaking scale can be lowered to $\sim1$~TeV without large weak-isospin violation, and a $126$~GeV Higgs boson mass can easily be obtained.  We also comment on the search of the new states at the LHC and future colliders.  The conclusions are drawn in Section~\ref{sec:con}. The two appendices collect the formula of $T$ parameter from fermion loops and the estimates of some model parameters.


\section{Composite Scalars with a Custodial Symmetry}
\label{sec:fullmodel}

As in the usual composite Higgs models, we assume that at a scale $\Lambda \gg 1$~TeV there are no fundamental scalars. To implement the custodial symmetry in the top seesaw dynamics, we introduce an $SU(2)_W$-singlet vector-like quark, $\chi$, of electric charge $+2/3$ and $SU(2)_W$-doublet vector-like quarks, $Q\equiv(X,T)$, with hypercharge $+7/6$,  in addition to the SM gauge group and fermions. For the doublet quarks, $T$ has electric charge $+2/3$, same as the SM top quark $t$, while $X$ has electric charge $+5/3$.  We assume that these new quarks, the left-handed  $(t_L, b_L)$ doublet and the right-handed $t_R$ in the SM (but not $b_R$) have some new non-confining strong interactions, which can be represented by 4-fermion interactions with strength proportional to  $1/\Lambda^2$. The strong dynamics is further assumed to approximately preserve the $U(5)_L \times U(4)_R$ chiral symmetry of the five left-handed fermions $\mathbf{\Psi}_L \equiv (t_L, b_L, X_L, T_L, \chi_L)$ and the four right-handed fermions $\mathbf{\Psi}_R \equiv (X_R, T_R, t_R, \chi_R)$.\footnote{The different orderings of left-handed and right-handed fermions are purely for convenience of later analysis.}  
The strong dynamics among the fermions at scale $\Lambda$ is given by
\begin{align}
\mathcal{L} =&~ \mathcal{L}_{\mbox{\scriptsize kinetic}}+G(\overline{\mathbf{\Psi}}_{L_i} \mathbf{\Psi}_{R_j})(\overline{\mathbf{\Psi}}_{R_j} \mathbf{\Psi}_{L_i}) \, .
\label{eq:lhe}
\end{align}

We assume the 4-fermion interactions in Eq.~(\ref{eq:lhe}) are sufficiently strong to form composite scalars that are quark-antiquark bound states.  These strong interactions are not confining, so that both the composite scalars and their constituents are present below the compositeness scale $\Lambda$.  The 4-fermion interactions give rise to the Yukawa couplings of the composite scalars (collectively labelled by $\Phi$) to their constituents and the masses of the scalars, 
\begin{align}
\mathcal{L}_{\mbox{\scriptsize Yukawa}}=-\xi \overline{\mathbf{\Psi}}_L\Phi \mathbf{\Psi}_R +\mbox{ H.c.} \, ,
\label{eq:Yukawa}
\\
\mathcal{L}_{\mbox{\scriptsize scalar masses}} = - M^2_{\Phi} \Tr[\Phi^\dagger \Phi] \, ,
\label{eq:scalarmass}
\end{align}
which also preserves the approximate $U(5)_L \times U(4)_R$ symmetry.  The scalar field $\Phi$ is a $5 \times 4$ complex matrix,
\begin{equation}
\Phi=
\begin{pmatrix}
\sigma^-_{tX} & \sigma^0_{tT} &  \sigma^0_{tt} & \phi^0_{t\chi}    \\
\sigma^{--}_{bX}  & \sigma^-_{bT}  &  \sigma^-_{bt} & \phi^-_{b\chi}    \\ 
 \sigma^0_{XX} & \sigma^+_{XT} & \sigma^+_{Xt} & \phi^+_{X\chi}  \\
 \sigma^-_{TX} &  \sigma^0_{TT} & \sigma^0_{Tt} & \phi^0_{T\chi}    \\ 
 \sigma^-_{\chi X} & \sigma^0_{\chi T} & \sigma^0_{\chi t} & \phi^0_{\chi\chi}
\end{pmatrix}
\equiv
\begin{pmatrix} \Sigma_X &\Sigma_T & \Sigma_t & \Phi_\chi \end{pmatrix}.
\label{eq:Phi}
\end{equation}
For each of the 20 complex scalars, the superscript denotes the electric charge and the subscript indicates the fermion constituents of the scalar.  For example, $\sigma^-_{tX} \sim ( t_L \overline{X}_R )$, and has electric charge $-1$.  The fields that contain $\overline{\chi}_R$ are labelled differently ($\phi$ instead of $\sigma$) because they contain the light scalars which will be the focus of our study.  It is useful to classify the scalar fields in Eq.~(\ref{eq:Phi}) into the following categories:
\begin{itemize}
  \item $\bpm \phi^0_{t\chi} \\  \phi^-_{b\chi} \epm$, $\bpm  \phi^+_{X\chi} \\  \phi^0_{T\chi} \epm$, 
  	  $\bpm \sigma^0_{tt} \\ \sigma^-_{bt} \epm$, $\bpm \sigma^+_{Xt} \\  \sigma^0_{Tt} \epm$, 
	  $\bpm \sigma^-_{\chi X} \\  \sigma^0_{\chi T} \epm$ are EW doublets;
  \item $\sigma^0_{\chi t}$ and $\phi^0_{\chi\chi}$ are EW singlets;
  \item $\begin{pmatrix}  \sigma^0_{XX} & \sigma^+_{XT} \\  \sigma^-_{TX} &  \sigma^0_{TT}   \end{pmatrix}$ contains one EW triplet and one singlet, which can be parameterized as
\begin{equation}
\begin{pmatrix} \sigma^+_{XT} \\ \frac{1}{\sqrt{2}}(\sigma^0_{XX} - \sigma^0_{TT}) \\ \sigma^-_{TX} \end{pmatrix} , ~~~~~~ \frac{1}{\sqrt{2}}(\sigma^0_{XX} + \sigma^0_{TT}),
\end{equation}
respectively.  Similarly, $\begin{pmatrix} \sigma^-_{tX} & \sigma^0_{tT}   \\ \sigma^{--}_{bX}  & \sigma^-_{bT}  \end{pmatrix}$ also contains one triplet $\bpm \sigma^0_{tT} \\ \frac{1}{\sqrt{2}}(\sigma^-_{tX} - \sigma^-_{bT}) \\ \sigma^{--}_{bX} \epm$ and one singlet $\frac{1}{\sqrt{2}}(\sigma^-_{tX} + \sigma^-_{bT})$.
\end{itemize}

The vector-like fermions can possess gauge invariant masses, which may be generated by the physics at some higher scale than $\Lambda$:
\begin{equation}
\mathcal{L}_{\mbox{\scriptsize fermion masses}} =
-\mu_t\overline{\chi}_Lt_R - \mu_{\chi\chi} \overline{\chi}_L\chi_R -  \mu_Q \begin{pmatrix} \overline{X}_L & \overline{T}_L \end{pmatrix} \begin{pmatrix} X_R \\ T_R \end{pmatrix} + \mbox{H.c.}
\label{eq:fermion_masses}
\end{equation}
These fermion mass terms explicitly break the $U(5)_L \times U(4)_R$ symmetry. They are assumed to be small compared to $\Lambda$ so that they do not affect the strong dynamics. Below the compositeness scale, these mass terms are matched to the tadpole terms of the composite scalars.

At scales $\mu<\Lambda$, the Yukawa couplings give rise to the quartic couplings and corrections to the masses of the scalars.  We assume that there are additional explicit $U(4)_R$ breaking effects which distinguish $t_R$, $\chi_R$ and $Q_R$.  Since mass terms are quadratically sensitive to the UV physics, such effects could induce a large relative splitting of the masses for $\Sigma_{X,T}$, $\Sigma_t$ and $\Phi_\chi$.  
Combining the quartic couplings, mass terms and tadpole terms, the scalar potential below scale $\Lambda$ is given by  
\begin{align}
V =&~ \frac{\lambda_1}{2}\Tr[(\Phi^\dagger \Phi)^2] +\frac{\lambda_2}{2}(\Tr[\Phi^\dagger \Phi])^2 \nonumber\\ 
& + M^2_{\Sigma_{X,T}}\Sigma_X^\dagger\Sigma_X + M^2_{\Sigma_{X,T}}\Sigma_T^\dagger\Sigma_T + M^2_{\Sigma_t}\Sigma_t^\dagger\Sigma_t  + M^2_{\Phi_\chi} \Phi_\chi^\dagger \Phi_\chi \nonumber\\
&-C_Q\sigma^0_{XX}-C_Q\sigma^0_{TT}-C_{\chi t}\sigma^0_{\chi t}-C_{\chi\chi}\phi^0_{\chi\chi} +\mbox{H.c.}
\label{eq:Vfull}
\end{align}
Because $Q_R\equiv(X_R,T_R)$ is an EW doublet, $\Sigma_X$, $\Sigma_T$ have the same mass-squared $M^2_{\Sigma_{X,T}}$, and $\sigma^0_{XX}$, $\sigma^0_{TT}$ have the same tadpole coefficient $C_Q$. (This guarantees that the VEV of triplet scalars are suppressed.)
Matching at the scale $\Lambda$, the size of the tadpole terms are related to the fermion mass terms by
\begin{equation}
C_Q \simeq \frac{\mu_Q}{\xi}\Lambda^2 ~, ~~~~ C_{\chi t} \simeq \frac{\mu_t}{\xi}\Lambda^2 ~, ~~~~ C_{\chi\chi} \simeq \frac{\mu_{\chi\chi}}{\xi}\Lambda^2 ~.
\end{equation}
When the scalars are integrated out at the cutoff scale, the fermion mass terms are recovered.  This means that at scale $\mu<\Lambda$ we do not need to include the explicit fermion mass terms in Eq.~(\ref{eq:lhe}). They will appear from the scalar VEVs in the low energy effective theory.
The quartic coupling $\lambda_1$ is generated by fermion loops and becomes non-perturbative near $\Lambda$.  $\lambda_2$, on the other hand, is not induced by fermion loops at the leading order and vanishes at $\Lambda$ in the large $N_c$ limit.  At scales $\mu<\Lambda$, scalar loops generate a non-zero value for $\lambda_2$ and give corrections to $\lambda_1$. Nevertheless, we expect $\lambda_1 \gg |\lambda_2|$.   
The spontaneous breaking of the chiral symmetry requires at least one of the scalars to have a negative mass-squared.  To obtain the correct SM limit, we require $M^2_{\Phi_\chi}<0$, while $M^2_{\Sigma_t}$ and $M^2_{\Sigma_{X,T}}$ are assumed to be positive for simplicity.

The theory below the compositeness scale $\Lambda$ is given by Eq.~(\ref{eq:Yukawa}) and Eq.~(\ref{eq:Vfull}).    Overall, the scalar sector contains 2 complex triplets, 5 complex doublets and 4 complex singlets.  The full theory is rather complicated.  However, the main focus of this paper is the low energy ($\mu\ll \Lambda$) phenomenology, in particular, the mass of the Higgs boson and the constraint from the weak-isospin violation $T$ parameter.   To produce the correct top seesaw mechanism, the SM Higgs doublet is required to be mostly the linear combination of $\bpm \phi^0_{t\chi} \\ \phi^-_{b\chi} \epm$ and $\bpm \phi^+_{X\chi} \\ \phi^0_{T\chi} \epm$, the doublet fields in $\Phi_\chi$.  Although a light $\Sigma_X$, $\Sigma_T$ or $\Sigma_t$ is not necessarily ruled out by current experimental constraints, from a naturalness point of view it is more reasonable to assume that their masses are not much smaller than $\Lambda$, so that all the degrees of freedom in them are heavy and can be integrated out for $\mu\ll \Lambda$ to obtain a low energy effective theory with $\Phi_\chi$ only.    We will focus on this low energy theory for the rest of this paper.


\section{Higgs Boson as a PNGB of the $O(5)$ Symmetry}
\label{sec:let}

We now study the effective theory at scale $\mu\ll \Lambda$ obtained by integrating out the heavy modes in $\Sigma_X$, $\Sigma_T$ and $\Sigma_t$. For simplicity we will sometimes label them collectively as $\Sigma_{X,T,t}$ and their masses as $M_{\Sigma_{X,T,t}}$.  In the effective theory, the lowest order contribution of $\Sigma_{X,T,t}$ simply comes from the VEVs of $\sigma^0_{XX}$, $\sigma^0_{TT}$ and $\sigma^0_{\chi t}$, induced by the tadpole terms in Eq.~(\ref{eq:Vfull}).  The subleading corrections, including the VEVs of other neutral fields in $\Sigma_{X,T,t}$, are suppressed by $1/M^2_{\Sigma_{X,T,t}}$.  We will first consider the contributions from VEVs of $\sigma^0_{XX}$, $\sigma^0_{TT}$ and $\sigma^0_{\chi t}$ only and study the $\mathcal{O}( 1/M^2_{\Sigma_{X,T,t}} )$ corrections later in Section~\ref{sec:mxt}.   

The scalar potential at $\mu\ll \Lambda$ can be written as
\begin{equation}
V = \frac{\lambda_1}{2}\Tr[(\Phi^\dagger \Phi)^2] +\frac{\lambda_2}{2}(\Tr[\Phi^\dagger \Phi])^2 + M^2_{\Phi_\chi} \Phi^\dagger_\chi \Phi_\chi -C_{\chi\chi}(\phi_\chi+\phi^\dagger_\chi),
\label{eq:Vl1}
\end{equation}
where\footnote{From now on, we will drop the subscript $\chi$ for the fields in $\Phi_\chi$ and sometimes the electric charge label as well (e.g.\ $\phi^0_{t\chi} \to \phi_t$) for convenience.}
\begin{equation}
\Phi =
\begin{pmatrix}
0 &0 &0 & \phi^0_t    \\
0 &0 &0 & \phi^-_b    \\
 \frac{w}{\sqrt{2}} &0 &0 & \phi^+_X  \\
 0& \frac{w}{\sqrt{2}} &0 & \phi^0_T    \\
 0&0 & \frac{u_t}{\sqrt{2}} & \phi^0_{\chi}
\end{pmatrix},~~~ 
\Phi_\chi =
\begin{pmatrix}
\phi^0_t    \\
 \phi^-_b    \\
 \phi^+_X  \\
 \phi^0_T    \\
 \phi^0_{\chi}
\end{pmatrix} ,  \label{eq:phi}
\end{equation}
$w$ and $u_t$ are defined as
\begin{equation}
\langle \sigma_{XX} \rangle = \langle \sigma_{TT} \rangle \equiv \frac{w}{\sqrt{2}} ~ , ~~~~~  \langle \sigma_{\chi t} \rangle \equiv \frac{u_t}{\sqrt{2}} ~ .
\label{eq:vev1}
\end{equation}
At the lowest order,  $\sigma_{XX}$ and $\sigma_{TT}$ have the same VEVs since they have the same tadpole terms.   This guarantees that the triplet scalar does not develop a VEV at the lowest order, which may otherwise cause a large weak isospin violation.  

Eq.~(\ref{eq:Vl1}) has an $U(5)_L$ chiral symmetry which is explicitly broken by  the heavy field VEVs $w$ and $u_t$ and the tadpole term $C_{\chi\chi}$.  Without the explicit breaking terms, $U(5)_L$ is spontaneously broken to $U(4)_L$ due to a negative mass-squared $M^2_{\Phi_\chi}$, and $\Phi_\chi$ contains 9 NGBs which includes two massless Higgs doublets.  If the explicit breaking is small, the theory will have two light Higgs doublet. Although the possibility of additional light scalars is not ruled out, such a theory will not have an EWSB minimum that approximately preserves the custodial symmetry. 
As we will see in Section~\ref{sec:fermion}, the VEV $w$ is constrained by the search of the charge $+5/3$ quark to be at least several hundred  GeV.  A large $w$ can raise the masses of one of the Higgs doublet by explicit breaking the $U(5)_L$ chrial symmetry down to an approximate $U(3)_L$ symmetry of $(\phi^0_t, \phi^-_b, \phi^0_\chi)$.  However, the $U(3)_L$ symmetry does not contain the $SU(2)$ custodial symmetry and we just recover the minimal model of Ref.~\cite{Cheng:2013qwa} in this limit,  which makes the extension of the hypercharge $+7/6$ quarks ($X$ and $T$) and the corresponding composite scalars totally pointless!
To solve this problem, we introduce the following mass terms (parameterized by the mass-squared parameter $K^2$) that also explicitly break $U(5)_L$,
\begin{equation}
V_{ {\rule[2pt]{16pt}{0.5pt}} \hspace{-16pt} U(5) }=\frac{1}{2}K^2\left( \Tr[\Sigma'^\dagger \Sigma']+ A^2_\chi \right),
\label{eq:K}
\end{equation}
where 
\begin{align}
\Sigma &\equiv \begin{pmatrix} \phi^{0*}_t & \phi^+_X \\ \phi^-_b & \phi^0_T \end{pmatrix} , \\
\Sigma' &\equiv  \frac{1}{\sqrt{2}}(\Sigma-\epsilon \Sigma^* \epsilon^T) = \frac{1}{\sqrt{2}}
\begin{pmatrix} \phi^{0*}_t-\phi^{0*}_T & \phi^+_X + \phi^+_b  \\ \phi^-_X + \phi^-_b  & -\phi^0_t+\phi^0_T \end{pmatrix},
\end{align}
and $A_\chi$ is the CP-odd field in $\phi^0_\chi$ shown later in Eq.~(\ref{eq:vha}).  
They can come from gauge invariant 4-fermion operators in the UV theory. We require $K^2$ to be positive.  Eq.~(\ref{eq:K}) lifts up the masses of $A_\chi$ and one linear combination of the two Higgs doublets. The $U(5)_L$ is broken down to $O(5)$, which transforms the four components of the remaining Higgs doublet and the real part of $\phi_\chi^0$.\footnote{Eq.~(\ref{eq:K}) actually preserves an additional $O(5)$ symmetry among the heavy Higgs doublet and $A_\chi$, which makes their mass degenerate in the limit of no other explicit symmetry breaking.  In principle we could assign a different mass term to $A_\chi$, making its mass a free parameter.  This is not so relevant in our model and for simplicity we will stick with Eq.~(\ref{eq:K}).}  The custodial symmetry will approximately hold as long as the value of $K^2$ is large enough ($K^2\gg \lambda_1 w^2$). (More explicitly discussion will be done in Section~\ref{sec:mh}.)
In this case, the theory has only 4 pNGBs that forms the light SM-like Higgs doublet from spontaneous breaking of $O(5)$ to $O(4)$. At the same time an approximate custodial symmetry is also retained.

Combining Eq.~(\ref{eq:Vl1}) and Eq.~(\ref{eq:K}), the scalar potential is
\begin{align}
V =&~ \frac{\lambda_1}{2}\Tr[(\Phi^\dagger \Phi)^2] +\frac{\lambda_2}{2}(\Tr[\Phi^\dagger \Phi])^2 + M^2_{\Phi_\chi} \Phi^\dagger_\chi \Phi_\chi \nonumber\\ 
&  + \frac{1}{2}K^2\left( \Tr[\Sigma'^\dagger \Sigma']+ A^2_\chi \right)-C_{\chi\chi}(\phi_\chi+\phi^\dagger_\chi) \, .
\label{eq:Vl2}
\end{align}
$\phi^0_t$, $\phi^0_T$ and $\phi^0_\chi$ develop VEVs from tadpoles, heavy field VEVs and the negative mass squared $M^2_{\Phi_\chi}$. We parameterize them as
\begin{equation}
 \phi^0_t  =\frac{v_t+h_t+iA_t}{\sqrt{2}} ~, \hspace{1cm}  \phi^0_T  =\frac{v_T+h_T+iA_T}{\sqrt{2}} ~, \hspace{1cm}  \phi^0_\chi  =\frac{u_\chi+h_\chi+iA_\chi}{\sqrt{2}} ~.
 \label{eq:vha}
\end{equation}
The electroweak VEV, $v = \sqrt{v^2_t+v^2_T}$, is required to be about $246$~GeV.  Due to the explicit breaking from the VEV $w$, $v_t > v_T$ is required for the potential to be at a minimum.  For convenience, we define the ratio $v_t/v_T$ as
\begin{equation}
\tan{\beta} \equiv \frac{v_t}{v_T} > 1 \, .
\end{equation}
$u_\chi$ is a singlet VEV which is expected to be 
significantly larger than the electroweak VEV.  The scale of $O(5)$ breaking is defined as\footnote{This is different from the definition of $f$ in Ref.~\cite{Cheng:2013qwa}, which was given by $f=\sqrt{v^2 + u^2_t + u^2_\chi}$.  $u_t$ and $w$ are induced by tadpoles so they mainly represent the explicit breaking instead of the spontaneous breaking of the chiral symmetry, so we exclude them in the definition of $f$, though it only makes little difference since $u_t^2, w^2 \ll u_\chi^2$. }
\begin{equation}
f=\sqrt{v^2_t + v^2_T + u^2_\chi} ~,
 \label{eq:f}
\end{equation}
which is conventionally called the chiral symmetry breaking scale.


\subsection{Extended top seesaw}
\label{sec:fermion}

Once the scalar fields develop VEVs as in Eq.~(\ref{eq:vev1})~and~(\ref{eq:vha}), the Yukawa couplings in Eq.~(\ref{eq:Yukawa}) generate the following mass terms of the fermions:    
\begin{equation}
\mathcal{L} \supset -\frac{\xi}{\sqrt{2}} \begin{pmatrix} \overline{t_L} & \overline{T_L} & \overline{\chi_L}  \end{pmatrix}
\begin{pmatrix} 0 & 0 & v_t \\ 0 & w & v_T \\ u_t & 0 & u_\chi  \end{pmatrix}
\begin{pmatrix} t_R \\ T_R \\ \chi_R \end{pmatrix}
-\frac{\xi w}{\sqrt{2}} \overline{X_L} X_R   ~.
\label{eq:fermass}
\end{equation}

The $X$ quark has electric charge $+5/3$ and does not mix with any other fermions.  Its mass is given by 
\begin{equation}
m_X = \frac{\xi w}{\sqrt{2}}~.
\label{eq:mx}
\end{equation}
The most recent CMS search has excluded the charge $+5/3$ quark with a mass below $800$~GeV at $95\%$ confidence level (CL), assuming that they decay exclusively to $tW$~\cite{Chatrchyan:2013wfa}.\footnote{This is a very good approximation in our model since the charged Higgs are much heavier than $X$ as shown in Section~\ref{sec:mh}.}  This constrains the value of $w$ to be at least a few hundred GeV.  The $T$ quark, on the other hand, mixes with $t$ and $\chi$ so that the $2\times 2$ mass matrix in the usual top seesaw model is extended to a $3\times 3$ mass matrix.  We denote the three mass eigenstates as $t_1$, $t_2$ and $t_3$, ordered by $m_{t_1} \leq m_{t_2} \leq m_{t_3}$.  Given that $w$ can not be too small ($w \gtrsim 300$~GeV for $\xi \sim 3.6$), the top quark is always the the lightest mass eigenstate $t_1$, and its mass ($m_{\mbox{\scriptsize top}} \equiv m_{t_1}$) is approximately given by
\begin{equation}
m^2_{\mbox{\scriptsize top}} \approx  \frac{\xi^2 v^2_t}{2} \frac{u^2_t}{f^2} ~.
\label{eq:mtop}
\end{equation}
As we will see later $f\gg w$ is required to obtain the correct Higgs mass. The lighter top-partner $t_2$ is mainly $T$, its mass $m_{t_2}$ is almost degenerate with $m_X$ due to the small mixing.  There is also a bound on $m_{t_2}$ from the searches of the heavy top-like quarks \cite{CMS:2013tda, TheATLAScollaboration:2013sha}, similar to but slightly weaker than the bound of $m_X$.  The heavier top partner $t_3$ is mostly the EW singlet $\chi$, with a mass given by $m_{t_3}\sim \xi f /\sqrt{2}$.
Finally, to obtain the correct top mass in Eq.~(\ref{eq:mtop}), we have the following constraint
\begin{equation}
\frac{u_t}{f} \approx \frac{y_t}{\xi \sin{\beta}}~,
\label{eq:utf}
\end{equation}
where $y_t$ is the SM top Yukawa coupling, define as $m^2_{\mbox{\scriptsize top}} \equiv \frac{y^2_t v^2}{2}\,$.\footnote{Since the mass of the heavier Higgs doublet in our model is much larger than $f$, it is not very useful to define $m^2_{\mbox{\scriptsize top}} \equiv \frac{y^2_t v^2_t}{2}$ as in 2HDM.}  

With the addition of the $X$ and $T$ quarks, $(t_L, b_L)$ and $(X_L, T_L)$  form a $(2,2)$ representation under $SU(2)_L \times SU(2)_R$, which contains the $SU(2)_C$ custodial symmetry after EWSB.  In the limit that the vector-like mass $\mu_Q$ vanishes (or equivalently $w=0$), there is no explicit violation of the custodial symmetry in the $(t_L, b_L, X_L, T_L)$ sector, which implies a negative $T$ parameter relative to the SM value because it removes the SM contribution from $(t_L, b_L)$. On the other hand, if $\mu_Q \to \infty$, then $(X, T)$ decouples and we recover the fermion sector of the minimal model, which gives a large positive contribution to $T$ if the chiral symmetry breaking scale is low. We expect that in a suitable range of the $X$, $T$ masses, the $T$ parameter can be small and consistent with the EW measurements. In Appendix~\ref{app:T}, we provide the full expression for the $T$-parameter calculated from fermion loops, which we use in the numerical calculations in Section~\ref{sec:plots}. Other contributions, such as the contribution from triplet scalar VEVs, are very small as long as the masses of heavy scalars $M_{\Sigma_{X,T,t}}$ are sufficiently large. In principle there could be additional model-dependent contributions from unknown UV physics. Here we assume that the custodial symmetry is a good symmetry in the UV and all major explicit breaking effects have been parameterized in our low energy effective theory, so that they are negligible. 

Since we only add vector-like quarks, the calculable contributions to the $S$ parameter is negligible. However, there could be UV contributions from heavy vector states \cite{Agashe:2004rs, Barbieri:2007bh, Marzocca:2012zn, Giudice:2007fh, Orgogozo:2012ct }.  While such contributions are model-dependent, they can be estimated to be \cite{Bellazzini:2014yua}
\begin{equation}
\hat{S} \sim \frac{m^2_W}{m^2_\rho} \, ,
\end{equation}
where $\hat{S} = g^2/(16\pi)\, S$ and $m_\rho$ is the mass scale of the heavy vector state.  
We expect such states to exist, as mentioned later in Secion~\ref{sec:ew},  which sets the scale where gauge-loop contributions are cut off.  For $m_\rho=3$~TeV, a typical value for $f\sim 1$~TeV, we have $S\sim0.08$, within the $68\%$ CL of the experimental constraint \cite{Baak:2012kk}.  A larger value of $S$ (up to $\sim 0.27$) may still be allowed if we arrange a larger value for $T$ as well, which can be easily achieved in this model.

The $Zb\bar{b}$ coupling has been a long-standing issue in beyond SM model building, particularly for composite Higgs models.  The measured value of the $Z\to b \bar{b}$ branching ratio ($R_b$)\cite{ALEPH:2005ab} was known to be larger than the SM prediction.  A recent calculation of $R_b$ including two-loops corrections \cite{Freitas:2012sy} suggests that the SM prediction for $Z\to b \bar{b}$ branching ratio ($R_b$) is $2.4\sigma$ smaller than the measured value~\cite{Baak:2012kk}.  On the other hand, the forward-backward asymmetry of the bottom quark $A^b_{FB}$ measured at the $Z$-pole exhibits a $2.5\sigma$ discrepancy with the SM prediction\cite{Baak:2012kk}.  The two notable discrepancies together prefer a larger $Zb_R\bar{b}_R$ coupling compared with the SM value and a $Zb_L\bar{b}_L$ coupling very close to the SM value \cite{Batell:2012ca}.

Our model, by construction, does not introduce any modification to the $Zb\bar{b}$ coupling at tree level.  However, there are corrections to $Zb_L\bar{b}_L$ at loop levels, since the custodial symmetry that protects the $Zb_L\bar{b}_L$ coupling is only approximately preserved by the new physics.  The scalars $\sigma^{--}_{bX}$, $\sigma^-_{bT}$, $\sigma^-_{bt}$ and $\phi^-_{b\chi}$ in Eq.~(\ref{eq:Phi}) couple $b_L$ with $X_R$, $T_R$, $t_R$ and $\chi_R$ respectively and will induce corrections to the $Zb_L\bar{b}_L$ coupling at one-loop level.  These corrections are suppressed, either by the large masses of the scalars or due to the vector-like nature of $X$, $T$ and $\chi$.\footnote{Using the results in Ref.\ \cite{Haber:1999zh}, we found that the corrections to the $Zb_L\bar{b}_L$ coupling from these vector-like quarks are proportional to $m^2_Z/m^2_{\mbox{\scriptsize fermion}}$.}  We found these corrections to be much smaller than the allowed deviation on the $Zb_L\bar{b}_L$ coupling.  Another contribution to the $Zb_L\bar{b}_L$ coupling comes from the mixing of the top with its vector-like partners.  The mixing between $t$ and $T$ is negligible in our model.  The correction due to the mixing between $t$ and $\chi$, though suppressed by $v^2/f^2$, could become non-negligible for small $f$.  Nevertheless, to fulfill the experimental constraints on the $Zb\bar{b}$ coupling, one needs to introduce additional new physics which enhances the $Zb_R\bar{b}_R$ coupling.  If $b_R$ also couples strongly to the new physics, it is possible to arrange it in some representation under the custodial symmetry that gives an significant enhancement to the $Zb_R\bar{b}_R$ coupling \cite{Agashe:2006at, Contino:2006qr, DaRold:2010as, Alvarez:2010js}.  
We will not discuss this possibility in this paper.


\subsection{Mass of the Higgs boson(s)}
\label{sec:mh}

Using the extremization conditions (requiring the linear terms of $h_t$, $h_T$, $h_\chi$ to vanish), one can write the dimensionful parameters $M^2_{\Phi_\chi}$, $K^2$ and $C_{\chi\chi}$ in the scalar potential in Eq.~(\ref{eq:Vl2}) in terms of the VEVs and quartic couplings,\footnote{To have $v_t^2,\, v_T^2 \ll f^2$, some tuning among $M^2_{\Phi_\chi}$, $K^2$, $C_{\chi\chi}$ and the explicit breaking VEVs $w$, $u_t$ induced by the tadpole terms is required. The required relation is quite complicated in general. Nevertheless, one can illustrate this tuning in the limit $w,\, \lambda_2 \to 0$. In this case, we have $-\frac{2M^2_{\Phi}}{\lambda_1} = f^2$ and $(\frac{2\sqrt{2}C_{\chi\chi}}{\lambda_1 u^2_t})^2=f^2-v^2$, suggesting that the second quantity need to be tuned slightly smaller than the first quantity to obtain $v^2 \ll f^2$.}
\begin{align}
M^2_{\Phi_\chi} &= - \frac{\lambda_1}{2}f^2- \frac{\lambda_2}{2}(f^2+u^2_t+2w^2)-\frac{\lambda_1 w^2 v_T}{2(v_t+v_T)} ~, \nonumber\\
K^2 &= \frac{\lambda_1 w^2 v_t v_T}{v^2_t-v^2_T}~,  \nonumber\\
C_{\chi\chi} &=  \frac{\lambda_1 u_\chi}{2\sqrt{2}}(u^2_t -\frac{w^2 v_T}{v_t+v_T} ) ~.
\label{eq:lin}
\end{align}
The second equation in Eq.~(\ref{eq:lin}) can be written as
\begin{equation}
\frac{K^2}{\lambda_1w^2} = \frac{\tan{\beta}}{\tan^2{\beta}-1} ~ ,
\end{equation}
which explicitly shows that $\tan{\beta}>1$ as $\frac{K^2}{\lambda_1w^2}$ is positive, and that the custodial symmetric limit $\tan\beta \to 1$ corresponds to $K^2\gg \lambda_1 w^2$.   
The constraint on the weak-isospin violation $T$ parameter puts an upper bound on $\tan\beta$. In Section~\ref{sec:plots} it will be shown that $\tan\beta$ can not be much larger than 1 if the chiral symmetry breaking scale is close to the weak scale ($f\sim1$~TeV). 

Substituting Eq.~(\ref{eq:lin}) back to the potential in Eq.~(\ref{eq:Vl2}), we can write the Higgs mass in terms of the VEVs and quartic couplings, which is the smallest eigenvalue of the $3\times3$ mass-squared matrix of the CP-even neutral scalars $(h_t, h_T, h_\chi)$.  It is useful to switch to the basis $(h_1, h_2, h_\chi)$ with the following rotation
\begin{equation}
\begin{pmatrix}
h_1    \\
h_2    \\
h_{\chi}
\end{pmatrix}
=
\begin{pmatrix}
\frac{v_t}{v} & \frac{v_T}{v} & 0 \\
-\frac{v_T}{v} & \frac{v_t}{v} & 0 \\
0 & 0 & 1
\end{pmatrix}
\begin{pmatrix}
h_t    \\
 h_T    \\
 h_{\chi}
\end{pmatrix},
\end{equation}
 where the electroweak VEV is purely associated with $h_1$.  In this basis, the mass-squared matrix is
 \begin{equation}
 \bpm
 (\lambda_1+\lambda_2)v^2 & 0 &  (\lambda_1+\lambda_2)u_\chi v \\
 0 & \frac{\lambda_1 w^2 v^2}{2(v^2_t-v^2_T)} & 0 \\
 (\lambda_1+\lambda_2)u_\chi v & 0 & (\lambda_1+\lambda_2) u^2_\chi +\frac{\lambda_1 u^2_t}{2} -\frac{\lambda_1 w^2 v_T}{2(v_t+v_T)}
 \epm ~.
 \label{eq:mmhhh}
 \end{equation}
.

One can see that in this basis $h_2$ is already a mass eigenstate. The $126$~GeV Higgs boson, on the other hand, should correspond to the lighter eigenstate of $(h_1,h_\chi)$.  At the leading order of $v^2/f^2$, the Higgs mass ($M_h$) is given by 
\begin{equation}
M_h^2 \approx (\lambda_1+\lambda_2) v^2 \frac{u^2_t - \frac{w^2v_T}{v_t+v_T}}{2f^2(1+\frac{\lambda_2}{\lambda_1})+u^2_t-\frac{w^2v_T}{v_t+v_T}} ~.
\label{eq:mh0}
\end{equation}
Since $\lambda_2$ is not generated by the fermion loops, we expect that $|\lambda_2/\lambda_1|\ll 1$. [The one-loop renormalization group (RG) estimate in Appendix~\ref{app:rge} gives $ -0.15 \lesssim \frac{\lambda_2}{\lambda_1} \lesssim 0$.] To obtain the correct top quark mass through the top seesaw mechanism, we need $u^2_t \ll f^2$. We also require  $u^2_t -\frac{w^2v_T}{v_t+v_T}>0$ for a positive Higgs mass-squared.  Using $0<u^2_t -\frac{w^2v_T}{v_t+v_T} \ll 2f^2(1+\frac{\lambda_2}{\lambda_1})$, Eq.~(\ref{eq:mh0}) can be simplified as
\begin{equation}
M_h^2 \approx \frac{\lambda_1v^2}{2f^2}(u^2_t-\frac{w^2v_T}{v_t+v_T}) ~.
\label{eq:mh01}
\end{equation}
Eq.~(\ref{eq:mh01}) also shows that the Higgs mass is independent of $\lambda_2$ at the leading order.  

One could see in Eq.~(\ref{eq:mh0}) and (\ref{eq:mh01}) that the Higgs mass is proportional to the combination $u^2_t-\frac{w^2v_T}{v_t+v_T}$, while $u_t$ and $w$ are VEVs that explicitly break the $SO(5)$ symmetry, as shown in Eq.~(3.10). (They are induced by the tadpole terms.) In the limit that the $SO(5)$ symmetry is exact, $u_t$ and $w$ vanish and the Higgs boson becomes massless, verifying its pNGB nature.  It is also interesting to note that $u_t$ and $w$ give opposite contributions to the Higgs boson mass. $u_t$ contributes to the singlet mass and makes it heavier than the doublet, therefore effectively makes the mass-squared of the Higgs field  more negative, resulting in a larger Higgs boson mass. On the other hand, $w$ contributes to the doublet mass and has the opposite effect.

Combining Eq.~(\ref{eq:mh01}) with Eq.~(\ref{eq:mx}) and Eq.~(\ref{eq:utf}), we obtain
\begin{equation}
M_h^2 \approx \frac{\lambda_1}{2\xi^2}( \frac{y^2_t}{\sin^2{\beta}}-\frac{m^2_X}{f^2}\frac{2}{1+\tan{\beta}}) v^2 ~,
\label{eq:mh02}
\end{equation}
where $\tan{\beta}\equiv v_t/v_T$, $y_t$ is the SM top Yukawa coupling 
and $m_X$ is the mass of the heavy quark with charge $+5/3$.  As mentioned earlier, for the case of small $f$ which we are interested in, $\tan{\beta}$ is restricted to be slightly larger than 1.  The correct Higgs mass ($126$~GeV) corresponds to $\lambda_h = M_h^2/v^2 \approx 0.26$ at the weak scale. It is typically obtained for 
\begin{equation}
m_X \lesssim f \, ,
\label{eq:mxf}
\end{equation}
based on this approximation.
The Higgs mass could be modified by the effects of EW interactions (Section~\ref{sec:ew}) and the masses of heavy scalars (Section~\ref{sec:mxt}).  However, Eq.~(\ref{eq:mxf})  still generally holds as shown later in Section~\ref{sec:plots} with numerical calculations.

For the other CP-even neutral scalars,
$h_2$ is already a mass eigenstate with mass-squared $\frac{\lambda_1 w^2 v^2}{2(v^2_t-v^2_T)}$.  Due to the $O(5)$ symmetry, the masses of the heavy doublet CP-even neutral ($h_2$), CP-odd neutral, and charged scalars all have the same mass at the lowest order, which we denote collectively as
\begin{equation}
M^2_H = \frac{\lambda_1 w^2 v^2}{2(v^2_t-v^2_T)} ~.
\label{eq:mh2}
\end{equation}
Comparing Eq.~(\ref{eq:mh2}) with Eq.~(\ref{eq:lin}), we notice that $M^2_H/ K^2>1$ and it approaches 1 as $K^2 \to \infty$.  A lower bound exists for $M_H$ by the fact $v_T^2>0$, 
\begin{equation}
M^2_H > \frac{\lambda_1 w^2}{2} = \frac{\lambda_1}{2\xi^2} 2 m^2_X ~.
\end{equation}
A large $K^2$, required if $f$ is small, would imply that these scalars are significantly heavier than the hypercharge $+7/6$ quarks, beyond any current experimental bounds. The heavier eigenstate of $(h_1, h_\chi)$ is mostly the EW singlet. Its mass-squared is approximately $(\lambda_1+\lambda_2) f^2$ which is also much larger than the current bounds.


\subsection{$O(5)$ breaking from electroweak interactions}
\label{sec:ew}

So far we have assumed that the mass and quartic terms in the potential respect the $O(5)$ symmetry and the only explicit $O(5)$ breaking comes from tadpole terms.  If there exist additional $O(5)$ breaking effects, they will feed into the mass and quartic terms through loops and affect the predictions of the model, such as the mass of the Higgs boson.  In our model, additional $O(5)$ breaking effects come from the SM $SU(2)_W\times U(1)_Y$ gauge interactions.  Since the couplings of SM gauge bosons to the two Higgs doublets are the same,  they actually preserve the $U(4)_L$ chiral symmetry of the $(\phi_t, \phi_b, \phi_X, \phi_T)$ scalars and do not generate different masses or quartic couplings for the two Higgs doublets.  Hence, the $O(5)$ breaking mass and quartic terms can be parameterized as
\begin{align}
\Delta V_{\mbox{\scriptsize breaking}} =&~ \frac{\kappa_1+\kappa_2}{2}(\phi^\dagger_t \phi_t+\phi^\dagger_b \phi_b+\phi^\dagger_X \phi_X+\phi^\dagger_T \phi_T)^2 \nonumber\\
&  +(\kappa'_1+\kappa'_2)(\phi^\dagger_t \phi_t+\phi^\dagger_b \phi_b+\phi^\dagger_X \phi_X+\phi^\dagger_T \phi_T)\phi^\dagger_\chi \phi_\chi \nonumber\\
& + \frac{\kappa'_1}{2}w^2(\phi^\dagger_X \phi_X+\phi^\dagger_T \phi_T) \nonumber\\
& + \kappa'_2(w^2+\frac{u^2_t}{2})(\phi^\dagger_t \phi_t+\phi^\dagger_b \phi_b+\phi^\dagger_X \phi_X+\phi^\dagger_T \phi_T) \nonumber\\
& +\Delta M^2 (\phi^\dagger_t \phi_t+\phi^\dagger_b \phi_b+\phi^\dagger_X \phi_X+\phi^\dagger_T \phi_T) ~,
\label{eq:dVew}
\end{align}
where we have assumed that the quartic terms are $U(4)_R$ symmetric for simplicity and parameterized the scalar fields as in Eq.~(\ref{eq:phi}).  Assuming that the $SU(2)_W\times U(1)_Y$ gauge interactions are the only $O(5)$ breaking contribution besides the tadpole terms, the parameters $\Delta M^2$ and $\kappa_{1(2)}$, $\kappa'_{1(2)}$ in Eq.~(\ref{eq:dVew}) are estimated to be
\begin{equation}
\Delta M^2 = \frac{3}{64\pi^2}(3g^2_2+g^2_1)M^2_\rho ~,
\label{eq:msew}
\end{equation}
and 
\begin{equation}
\frac{\kappa_{1(2)}}{\lambda_{1(2)}} \simeq 2 \frac{\kappa'_{1(2)}}{\lambda_{1(2)}} \simeq
 \frac{3}{16\pi^2}(3g^2_2+g^2_1)\, \log{\frac{M_\rho}{\mu}} ~,
 \label{eq:lambdaew}
\end{equation}
where $g_2$ and $g_1$ are the $SU(2)_W\times U(1)_Y$ gauge couplings.  The cutoff  of the divergent integrals should be set by the mass of some strongly interacting (presumably vector) state in this theory. It could be $m_\rho$ discussed in Section~\ref{sec:fermion} up to an ${\cal O}(1)$ factor. We parameterize this cutoff by $M_\rho$.

It is straightforward to repeat the analysis in Section~\ref{sec:mh} by including Eq.~(\ref{eq:dVew}).  Keeping the leading order in $(v^2,u^2_t, w^2)/f^2$, $\Delta M^2$, $\kappa_1$ and $\kappa'_1$, the correction to $M^2_h$ is
 \begin{equation}
\Delta M^2_h \simeq \left(\kappa_{12} - \frac{5}{2}\kappa'_{12} - \frac{\Delta M^2}{f^2} \right) v^2~,
\label{eq:mhew}
 \end{equation}
 where $\kappa_{12} \equiv \kappa_1+\kappa_2$ and $\kappa'_{12} \equiv \kappa'_1+\kappa'_2$.  Since $\kappa_{12} \simeq 2\kappa_{12}$, we conclude that the contribution from EW interactions always decrease the Higgs mass.

Additional $O(5)$ breaking effects may exist besides the $SU(2)_W\times U(1)_Y$ gauge interactions.  In principle, these effects could break the $U(4)_L$ symmetry, but in order to avoid large violation of custodial symmetry, they should at least approximately preserve $O(4)$.  If the $U(4)_L$ breaking effects are mainly in the mass term, it effectively causes a shift of the $K^2$ terms in Eq.~(\ref{eq:Vl2}) except for $A^2_\chi$, and results in a splitting between the mass of $A^2_\chi$ and the mass of the heavy Higgs doublet.


\subsection{Corrections from heavy scalars masses}
\label{sec:mxt}

\begin{figure}
\centering
\includegraphics[trim = 6cm 19.5cm 6.5cm 3.5cm, clip, width=10cm]{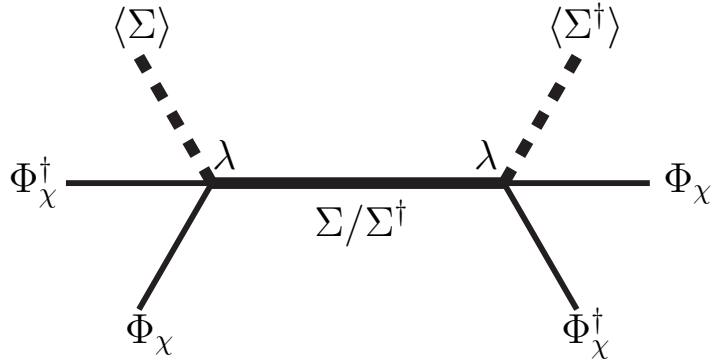}
\caption{The tree-level diagram which corresponds to the dimensional-six operators of the form $\frac{\lambda^2}{M^2}\Sigma^\dagger \Sigma \Phi^\dagger_\chi \Phi_\chi \Phi^\dagger_\chi \Phi_\chi$.  The thin lines represent $\Phi_\chi$, the thick line represents the heavy field $\Sigma$, and the thick dash lines are the heavy field VEVs $\langle \Sigma \rangle$ (i.e.\  $\langle \sigma^0_{XX} \rangle$, $\langle \sigma^0_{TT} \rangle$ or $\langle \sigma^0_{\chi t} \rangle$).}
\label{fig:d6}
\end{figure}

In Sec.~\ref{sec:mh} we have only included the lowest order contributions from heavy scalar fields $\Sigma_{X,T,t}$, which are the VEVs of $\sigma^0_{XX}$, $\sigma^0_{TT}$ and $\sigma^0_{\chi t}$.  We now study the corrections that are proportional to $1/M^2_{\Sigma_{X,T,t}}$.  As long as $M^2_{\Sigma_{X,T,t}}$ are large, $\Sigma_{X,T,t}$ can be integrated out and the dominate contributions come from the dimension-six operators of the form $\frac{\lambda^2}{M^2}\Sigma^\dagger \Sigma \Phi^\dagger_\chi \Phi_\chi \Phi^\dagger_\chi \Phi_\chi$. They are generated by the tree-level diagram in Fig.~\ref{fig:d6}, where we use $\Sigma$ and $\lambda$ to denote a general heavy field and a general quartic coupling.  Replacing the heavy fields with their VEVs, these operators generate quartic couplings of the $\Phi_\chi$ fields that explicitly break $O(5)$ and hence modify the Higgs mass.

With the quartic couplings in Eq.~(\ref{eq:Vl2}), we can write down the terms generated by Fig.~\ref{fig:d6}.  For simplicity, we assume all the fields in $\Sigma_X$ and $\Sigma_T$ have mass $M_{\Sigma_{X,T}}$ and all the fields in $\Sigma_t$ has mass $M_{\Sigma_t}$, which is an good approximation for large $M_{\Sigma_{X,T,t}}$ where the corrections from the tadpoles of $\sigma^0_{XX}$, $\sigma^0_{TT}$ and $\sigma^0_{\chi t}$ are negligible.  For simplicity, we also ignore the contributions from EW interactions discussed in Section~\ref{sec:ew}.  (At the lowest order, different contributions add up linearly.)  Thus, the leading correction from heavy scalars masses to the scalar potential is
\begin{align}
\Delta V =\,& -\frac{w^2(\lambda_1\phi^\dagger_X\phi_X + \lambda_2\Phi^\dagger_\chi \Phi_\chi)^2}{2M^2_{\Sigma_{X,T}}}  
-\frac{w^2\lambda^2_1}{2M^2_{\Sigma_{X,T}}}(\phi^\dagger_X\phi_X\Phi^\dagger_\chi\Phi_\chi - \phi^\dagger_X\phi_X\phi^\dagger_X\phi_X)
 \nonumber\\
& -\frac{w^2(\lambda_1\phi^\dagger_T\phi_T + \lambda_2\Phi^\dagger_\chi \Phi_\chi)^2}{2M^2_{\Sigma_{X,T}}}
-\frac{w^2\lambda^2_1}{2M^2_{\Sigma_{X,T}}}(\phi^\dagger_T\phi_T\Phi^\dagger_\chi\Phi_\chi - \phi^\dagger_T\phi_T\phi^\dagger_T\phi_T)
\nonumber\\
&-\frac{u^2_t(\lambda_1\phi^\dagger_\chi\phi_\chi + \lambda_2\Phi^\dagger_\chi \Phi_\chi)^2}{2M^2_{\Sigma_t}}
-\frac{u^2_t\lambda^2_1}{2M^2_{\Sigma_t}}(\phi^\dagger_t\phi_t\Phi^\dagger_\chi\Phi_\chi - \phi^\dagger_t\phi_t\phi^\dagger_t\phi_t) ~.
\label{eq:dVhs}
\end{align}
In the limit $\lambda_2\to 0$, the above expression is simplified to
\begin{equation}
\Delta V = -\frac{\lambda^2_1 w^2}{2M^2_{\Sigma_{X,T}}}(\phi^\dagger_X \phi_X + \phi^\dagger_T \phi_T)\Phi^\dagger_\chi \Phi_\chi - \frac{\lambda^2_1 u^2_t}{2M^2_{\Sigma_t}}\phi^\dagger_\chi \phi_\chi \Phi^\dagger_\chi \Phi_\chi ~.
\label{eq:dVsimp}
\end{equation}

Again, it is straightforward to calculate the effects of Eq.~(\ref{eq:dVhs}) on the Higgs mass by repeating the analysis in Section~\ref{sec:mh}.    For simplicity we set $\lambda_2=0$.  Keeping the lowest orders in terms of ${1}/{M^2_{\Sigma_{X,T}}}$ and ${1}/{M^2_{\Sigma_t}}$, we have
\begin{equation}
M_h^2 \approx \frac{\lambda_1v^2}{2f^2}\left[ u^2_t(1-\frac{\lambda_1f^2}{2M^2_{\Sigma_t}}) -\frac{w^2v_T}{v_t+v_T}(1-\frac{\lambda_1f^2}{2M^2_{\Sigma_{X,T}}})  \right] \, .
\label{eq:mhx03}
\end{equation}
Compared with Eq.~(\ref{eq:mh01}), we find that the Higgs mass $M_h$ decreases as $M_{\Sigma_t}$ decreases or $M_{\Sigma_{X,T}}$ increases, and vice versa.

The other contribution comes from the VEVs of the other neutral components of $\Sigma_{X,T,t}$, which are  $\sigma^0_{tT}$, $\sigma^0_{\chi T}$, $\sigma^0_{tt}$ and $\sigma^0_{\chi t}$ in Eq.~(\ref{eq:Phi}).  These fields do not have tadpole terms generated by gauge invariant fermion masses.  However, once other fields develop VEVs, the quartic couplings will induce  VEVs for these fields that are suppressed by $1/M^2_{\Sigma_{X,T,t}}$.  Compared to the leading order corrections in Eq.~(\ref{eq:mhx03}) that are proportional to $\frac{\lambda_1 f^2}{2M^2}$, the effects coming from these quartic-coupling-induced VEVs are further suppressed by at least a factor of $w^2/f^2$ or $v^2/f^2$.  The contribution to $S$ and $T$ parameters from the  triplet scalar VEVs are also negligible as long as $M^2_{\Sigma_{X,T}}$ is significantly large.  We will ignore these effects for simplicity.


\section{Numerical Studies and Phenomenology}    
\label{sec:plots}

In this section, we perform numerical studies of this model to obtain predictions and preferred ranges of the parameters, given the experimental constraints. They serve to verify the approximate analytic results obtained in the previous sections.  We also discuss possible phenomenologies at the LHC or future colliders.

We start with an enumeration of the parameters of this model.  At energy scale $\mu\ll \Lambda$, the theory is described by the scalar potential Eq.~(\ref{eq:Vl2}), where the composite scalars have the forms in Eq.~(\ref{eq:phi}), with the corrections from EW interactions in Eq.~(\ref{eq:dVew}), (\ref{eq:msew}) and (\ref{eq:lambdaew}), and the effects of heavy scalar masses in Eq.~(\ref{eq:dVhs}).  Together with the Yukawa sector, the theory has the following set of parameters,
\begin{equation}
\xi \,,~ \lambda_1 \,,~ \lambda_2 \,,~ M^2_{\Phi_\chi} \,,~ K^2 \,,~ C_{\chi\chi} \,,~ M_{\Sigma_t} \,,~ M_{\Sigma_{X,T}} \,,~ M_\rho \,,~ w \,,~ u_t~.
\end{equation}
Using Eq.~(\ref{eq:lin}), $M^2_{\Phi_\chi}$, $K^2$ and $C_{\chi\chi} $ can be written in terms of the VEVs $v$, $f$, $\tan{\beta}(\equiv v_t/v_T)$ and other parameters, where $v$ is fixed by the EW scale.  To produce the correct $m_{\mbox{\scriptsize top}}$, we use the SM 1-loop RG equations to evolve the SM top Yukawa coupling $y_t$ to the scale of the heavier top partner, $m_{t_3}\sim \xi f/\sqrt{2}$, and use it to solve for $u_t$, which in the lowest order is given by Eq.~(\ref{eq:utf}).  The running top Yukawa coupling in the $\overline{\rm MS}$ scheme at the scale $m_{\mbox{\scriptsize top}}$ corresponds to $m_{\mbox{\scriptsize top}} (\mu = m_{\mbox{\scriptsize top}})\approx 160$~GeV~\cite{Langenfeld:2009wd}.  $w$ is related to the mass of the charge $+5/3$ quark $m_X$ by $m_X = \xi w / \sqrt{2}$.  Hence, the spectrum is fully determined by the following parameters,
\begin{equation}
\xi \,,~ \lambda_1/(2\xi^2) \,,~ \lambda_2/\lambda_1 \,,~ f \,,~ \tan{\beta} \,,~  M_{\Sigma_t} \,,~ M_{\Sigma_{X,T}} \,,~ M_\rho \,,~ m_X~.
\label{eq:para}
\end{equation}
We choose the ratios of couplings $\lambda_1/(2\xi^2)$ and $\lambda_2/\lambda_1$ as the independent parameters because they are more convenient and better constrained.  To calculate $M_h$, we match the theory to the SM at the scale of the heavier top partner $m_{t_3}$, compute the quartic Higgs coupling $\lambda_h$, and then evolve $\lambda_h$ down to the weak scale.

Before starting the numerical calculations, we first examine the expected ranges of input parameters listed in Eq.~(\ref{eq:para}).  The Yukawa coupling $\xi$ is expected to be $\sim3-4$ in a strongly coupled theory.  We will use $\xi \approx 3.6$ as the standard reference value~\cite{Cheng:2013qwa}.  The ranges of $\lambda_1/(2\xi^2) \,,~ \lambda_2/\lambda_1$ are discussed in Appendix~\ref{app:rge} and are expected to be $0.35 \lesssim \lambda_1/(2 \xi^2) \lesssim 1$ and  $-0.15 \lesssim \lambda_2/\lambda_1 \lesssim 0$.  Since the focus of this paper is to reduce the chiral symmetry breaking scale $f$ without violation of experimental constraints, we will consider lower values of $f$ ($ \lesssim 5$~TeV).  We often take $f=1$~TeV as a benchmark point. As we will see later, to obtain a correct Higgs mass $f$ can not be much smaller than $1$~TeV.  In Section~\ref{sec:let}, we already saw that $\tan{\beta}>1$ is required for the potential to be at a minimum.\footnote{A slightly larger lower bound on $\tan{\beta}$ can be obtained by imposing $M_H$ in Eq.~(\ref{eq:mh2}) to be smaller than the compositeness scale, which at most gives $\tan{\beta}\gtrsim 1.01$ and is irrelevant for our study. }  For small $f$, we expect $\tan{\beta}$ to be not much larger than 1 from the $T$ parameter constraint.  For the effective theory below the composite scale $\Lambda$ to be a valid description, the states in the theory should have masses below $\Lambda \sim 4\pi f$.  Furthermore, for the effective theory at $\mu\ll \Lambda$ described in Section~\ref{sec:let} to be a valid description, the heavy scalar masses $M_{\Sigma_{X,T,t}}$ need to be much larger than $f$.  Thus, we require $M_\rho\lesssim 4\pi f$ and $f \ll M_{\Sigma_{X,T,t}}\lesssim 4\pi f$.  Finally, the current bound from LHC requires $m_X>0.8$~TeV.

\begin{figure}[!]
\centering
\includegraphics[height=8cm]{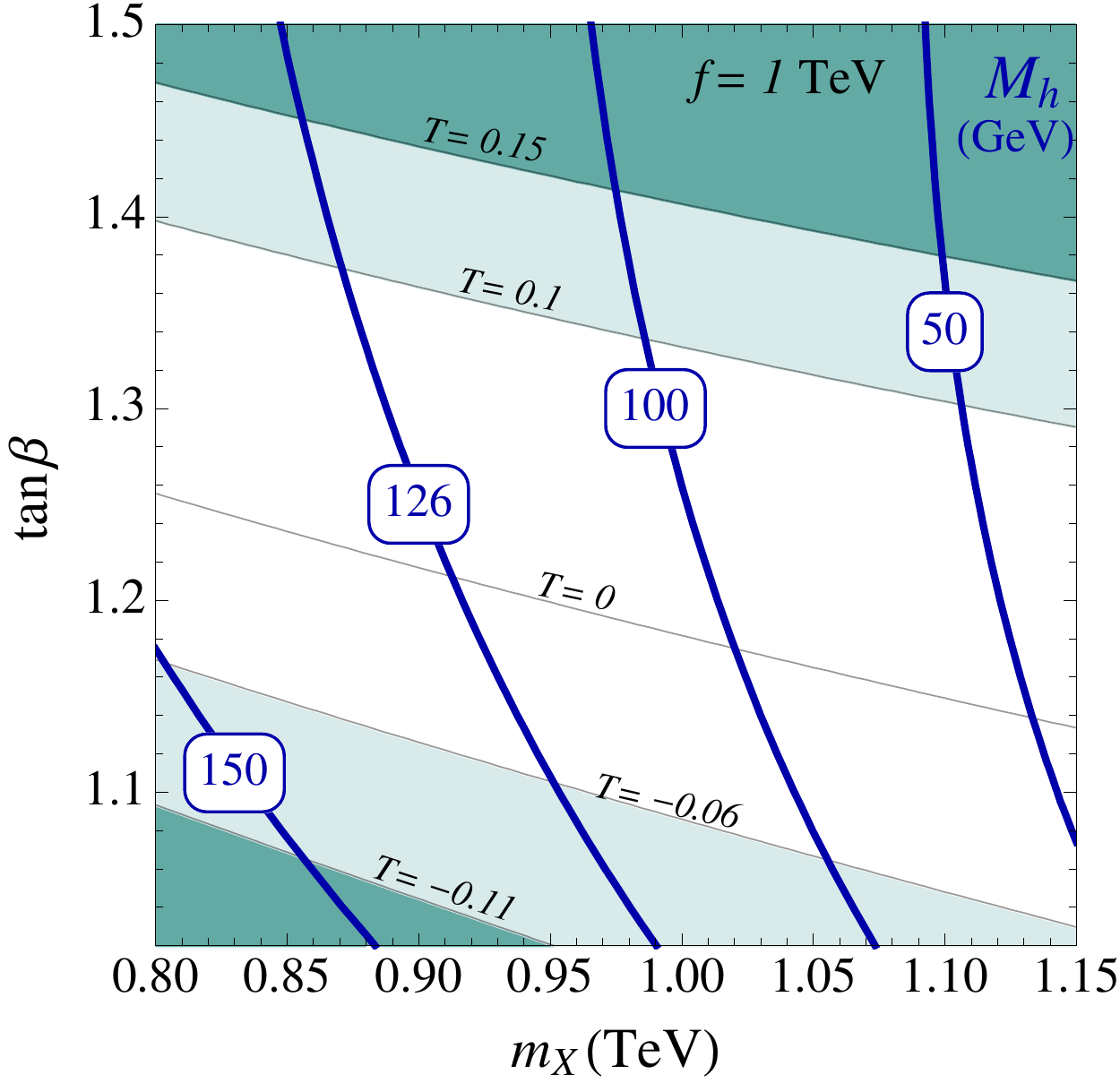}
\caption{Higgs boson mass as a function of $m_X$ and $\tan{\beta}$.   We fix $f=1$~TeV and choose the other parameters to be $\xi=3.6$, $\lambda_1/(2\xi^2)=0.7$, $\lambda_2/\lambda_1=0$, $M_\rho=3f$ and $M_{\Sigma_{X,T,t}}=10 f$.  The $68\%$ and $95\%$ CL for the $T$ parameter roughly corresponds to $-0.06<T<0.1$ and $-0.11<T<0.15$ (fixing $S=0$), which are shown on the plots with different color regions.}
\label{fig:mh1}
\end{figure}

In this model, we incorporate the custodial symmetry by introducing a vector-like EW doublet $(X,T)$, in order to reduce the chiral symmetry scale $f$ without introducing large weak isospin violation.  We first would like to verify whether this can indeed be achieved.  In Fig.~\ref{fig:mh1}, we show the Higgs boson mass $M_h$ as a function of $m_X$ and $\tan{\beta}$, by fixing $f=1$~TeV and other parameters to some typical values, $\xi=3.6$, $\lambda_1/(2\xi^2)=0.7$, $\lambda_2/\lambda_1=0$, $M_\rho=3f$.
For simplicity, we set the heavy scalar masses to be $M_{\Sigma_{X,T,t}}=10 f$, a value close to the compositeness scale.   We also show the contours of the $T$ parameter calculated using the expressions in Appendix~\ref{app:T}.  The regions  $-0.06<T<0.1$ and $-0.11<T<0.15$ roughly correspond to the $68\%$ and $95\%$ CL (fixing $S=0$) \cite{Baak:2012kk}, which are shown on the plots with different colors.  We see that, indeed, there is a region for which the $T$ parameter is within the constraint, while a $126$~GeV Higgs boson mass can also be obtained.  This demonstrates that the chiral symmetry breaking scale can be lowered from multi-TeV in the minimal model~\cite{Cheng:2013qwa} to $\sim 1$~TeV, which greatly reduces the tuning.  In Section~\ref{sec:mh} we argued that with small $f$, $\tan{\beta}$ can not be much larger than 1, as otherwise the custodial symmetry is badly broken.  This is verified in Fig.~\ref{fig:mh1}, as one can see the $68\%$ CL bound of the $T$ parameter requires $\tan{\beta}<1.4$.  On the other hand, a small custodial breaking is needed to account for the $(t_L, b_L)$ contribution in the SM, which translates into a lower bound on $\tan\beta$ when $m_X$ is small.

\begin{figure}[t!]
\centering
\includegraphics[height=7cm]{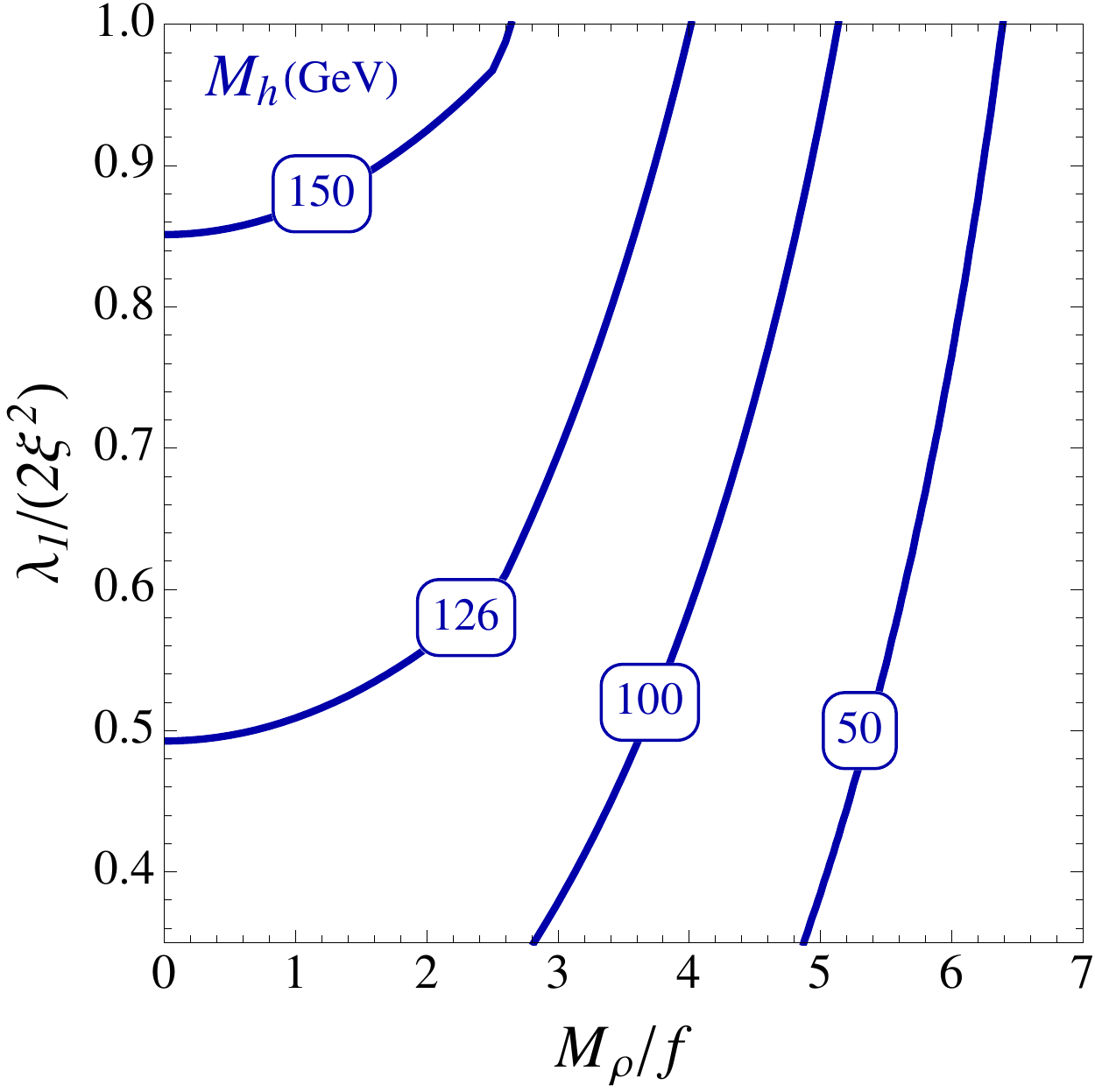} \hspace{0.5cm}
\includegraphics[height=7cm]{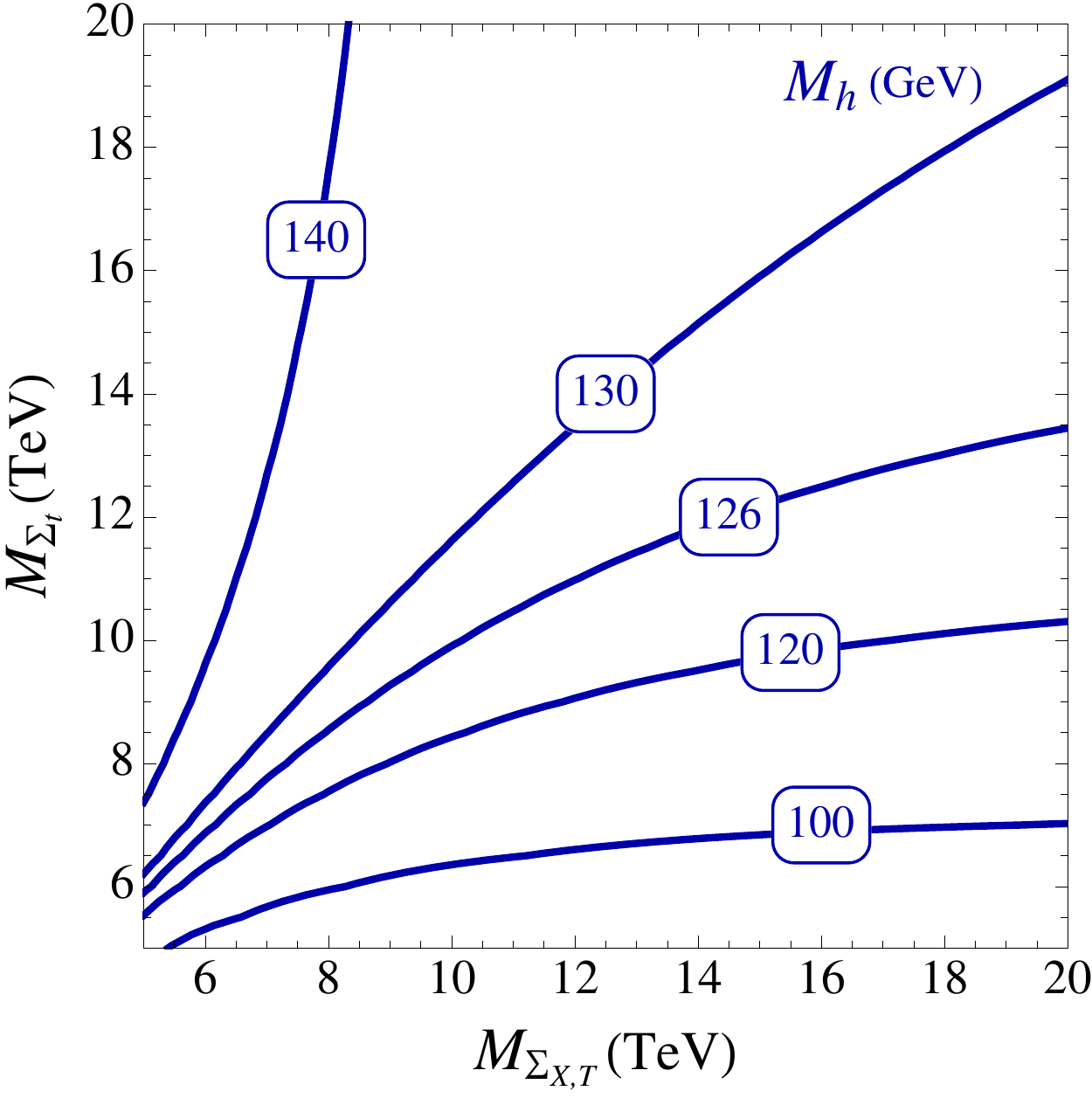}
\caption{{\bf Left:} Higgs boson mass as a function of $M_\rho/f$ and $\lambda_1/(2\xi^2)$.   {\bf Right:} Higgs boson mass as a function of $M_{\Sigma_{X,T}}$ and $M_{\Sigma_t}$.   For both plots, we set $\xi=3.6$, $\lambda_2/\lambda_1=0$, $f=1$~TeV, $\tan{\beta}=1.25$ and $m_X=0.9$~TeV.  In the plot on the left, we fix $M_{\Sigma_{X,T,t}}=10 f$; for the one on the right, we fix $\lambda_1/(2\xi^2)=0.7$ and $M_\rho=3f$.  The $T$ parameter is 0.02 for both plots.}
\label{fig:mh2}
\end{figure}

The Higgs boson mass is sensitive to $\lambda_1/(2\xi^2)$ and $M_\rho/f$.  To study the effects of these two parameters, we choose a point in Fig.~\ref{fig:mh1}, $m_X=0.9$~TeV and $\tan{\beta}=1.25$, then vary $\lambda_1/(2\xi^2)$ and $M_\rho/f$ and plot the Higgs boson mass as a function of these two parameters.  The result is shown in the left panel of Fig.~\ref{fig:mh2}.  Due to the running effects, the Higgs boson mass-squared does not vary linearly with $\lambda_1/(2\xi^2)$ as na\"{\i}vely indicated from Eq.~(\ref{eq:mh02}). The dependence is somewhat less sensitive.  The Higgs mass decreases as one increases $M_\rho$ as expected from Eq.~(\ref{eq:mhew}).  If $M_\rho$ is not too large ($M_\rho\lesssim 7f$), its effect can be compensated by different choices of other parameters to obtain the correct Higgs mass.

The Higgs boson mass also receives corrections from the masses of heavy scalars $\Sigma_{X,T,t}$.  We repeat the same exercise (choosing the point in Fig.~\ref{fig:mh1} with $m_X=0.9$~TeV and $\tan{\beta}=1.25$) and plot the Higgs boson mass as a function of $M_{\Sigma_{X,T}}$ and $M_{\Sigma_t}$.
A larger $M_h$ occurs for a larger $M_{\Sigma_t}$ and a smaller $M_{\Sigma_{X,T}}$, which agrees with the approximate formula Eq.~(\ref{eq:mhx03}) in Section~\ref{sec:mxt}.  If $M_{\Sigma_{X,T}}$ and $M_{\Sigma_t}$ are fixed to be the same ($M_{\Sigma_{X,T}}=M_{\Sigma_t}=M_{\Sigma_{X,T,t}}$), we see that the Higgs mass is not very sensitive to $M_{\Sigma_{X,T,t}}$, decreasing only by $\sim10$~GeV for $M_{\Sigma_{X,T,t}}$ going from $20$~TeV to $6$~TeV.  It is possible to  introduce large corrections to $M_h$ by arranging a large hierarchy between $M_{\Sigma_{X,T}}$ and $M_{\Sigma_t}$, but it is unnatural for either of them to be much smaller than the compositeness scale.  A very small $M_{\Sigma_{X,T}}$ or $M_{\Sigma_t}$ also makes the effective theory approach in Section~\ref{sec:mxt} unjustified.  The fact that $M_h$ is not very sensitive to $M_{\Sigma_{X,T}}$ and $M_{\Sigma_t}$ within reasonable ranges of the two parameters justifies our choice of $M_{\Sigma_{X,T}} = M_{\Sigma_t}=10 f$ in Fig.~\ref{fig:mh1}.  For simplicity, we also fix $M_{\Sigma_{X,T}} = M_{\Sigma_t}=10 f$ for the other plots in this section.

\begin{figure}[t!]
\centering
\includegraphics[height=7.4cm]{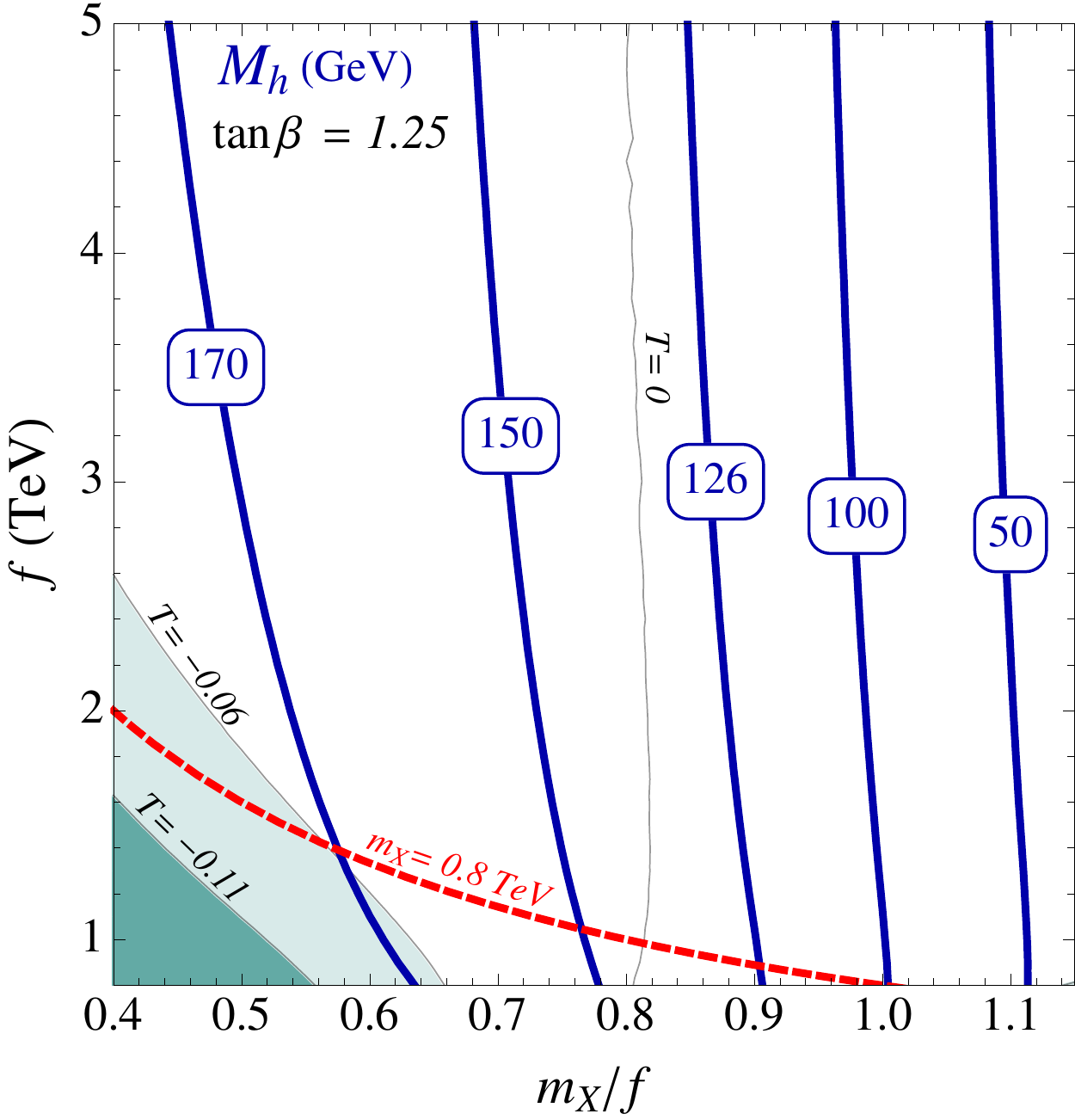} \hspace{0.5cm}
\includegraphics[height=8.2cm]{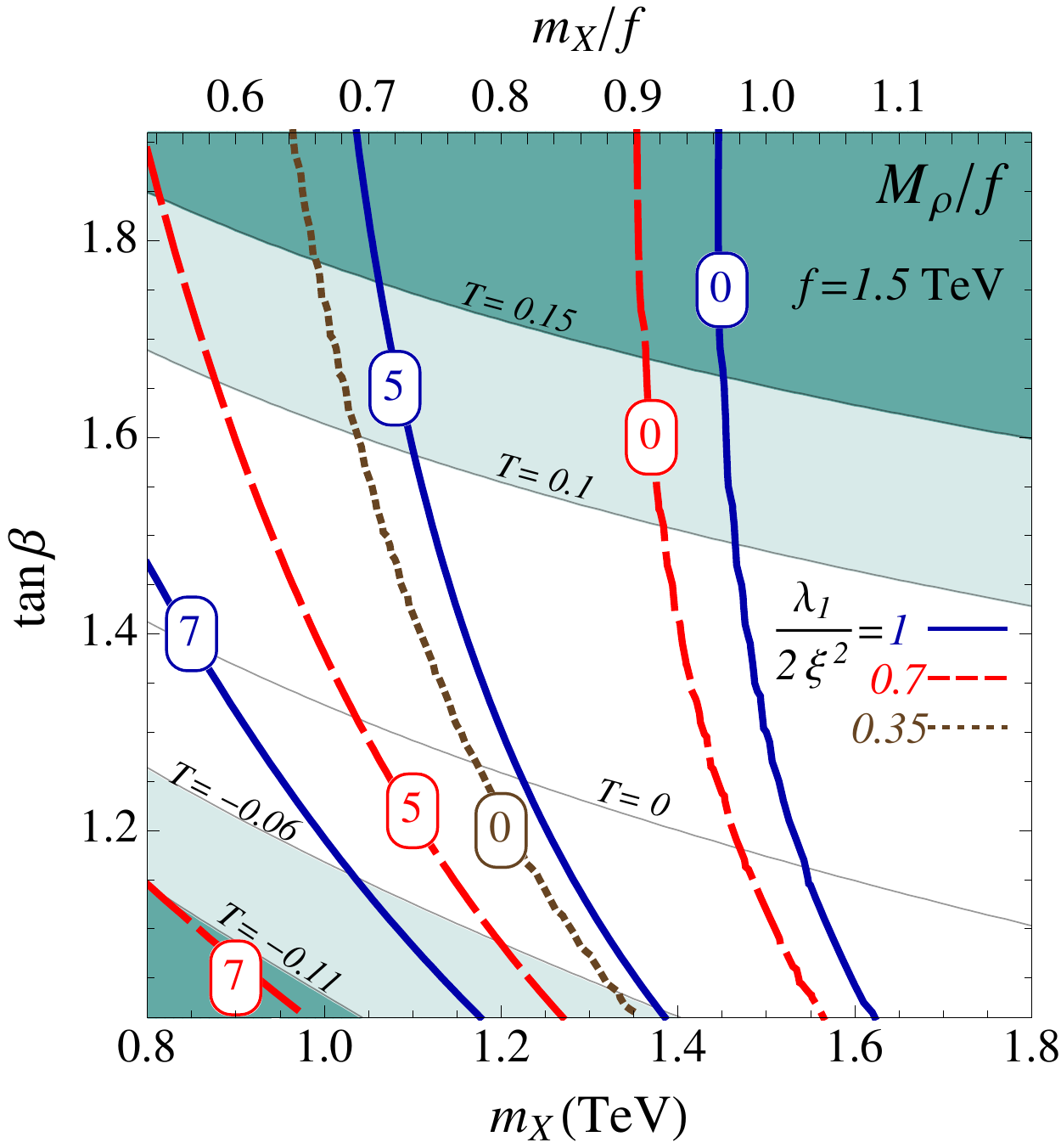} 
\caption{ {\bf Left:} Higgs boson mass as a function of $m_X/f$ and $f$.  We set the other parameters to be $\xi=3.6$, $\lambda_1/(2\xi^2)=0.7$, $\lambda_2/\lambda_1=0$, $\tan{\beta}=1.25$, $m_X=0.9$~TeV, $M_\rho=3f$ and $M_{\Sigma_{X,T,t}}=10 f$.  {\bf Right:} Contour plots of $M_\rho/f$ in the $(m_X, \tan{\beta})$ plane with the Higgs boson mass fixed at $126$~GeV.   We fix $f=1.5$~TeV and other parameters are set to be $\xi=3.6$, $\lambda_2/\lambda_1=0$, $M_{\Sigma_{X,T,t}}=10 f$.}
\label{fig:mh3}
\end{figure}

Eq.~(\ref{eq:mh02}) suggests that the Higgs boson mass is sensitive to the ratio $m_X/f$ rather than the individual value of $m_X$ or $f$.  This is verified in Fig.~\ref{fig:mh3} (left panel), where we show the Higgs mass as a function of $m_X/f$ and $f$ while fixing the other parameters to some typical values.  
We see the contours of constant Higgs masses are almost vertical unless $m_X/f$ is very small.  This suggests that the mass of the heavier top partner, $m_{t_3}\sim \xi f /\sqrt{2}$, is approximately proportional to $m_X$ (and $m_{t_2}$ since $m_{t_2} \approx m_X$) if the Higgs boson mass and other parameters are fixed. This is different from the predictions of many other composite Higgs models that contain 
more than one top partners, such as MCHM$_5$ and MCHM$_{10}$ in Ref.~\cite{Contino:2006qr, Pomarol:2012qf}.  In practice, the required ratio $m_X/f$ depends on other parameters that affect the Higgs boson mass, such as $\lambda_1/(2\xi^2)$ and $M_\rho$, which are not known {\it a priori}.  Nevertheless, for any reasonable set of other parameters, we could find the corresponding value of $m_X/f$ to give $M_h=126$~GeV.
In the right panel of Fig.~\ref{fig:mh3}, we fix $f=1.5$~TeV and use the constraint $M_h=126$~GeV to determine $M_\rho$ for different points in the $(m_X, \tan{\beta})$ plane and plot the value of $M_\rho$ in that plane.  In the plot we show three sets of contours which correspond to $\lambda_1/(2\xi^2) = 1,\, 0.7, \, 0.35\,$, covering the expected range $0.35 \lesssim \lambda_1/(2 \xi^2) \lesssim 1 \,$.  The $M_\rho=0$ contours represent the case where the explicit $O(5)$ breaking from the EW gauge loops is absent.  A larger $m_X/f$ reduces the Higgs mass, so the maximum value of $m_X/f$ occurs for the largest possible $\lambda_1/(2\xi^2)(=1)$ and the smallest $M_\rho$, for $M_h$ fixed at 126~GeV.  This is indeed the case in Fig.~\ref{fig:mh3}, and we find that the upper bound of $m_X/f$ is around $1$.  We have also verified numerically that for different values of $f$, we always have $m_X/f \lesssim 1$.  On the other hand, by choosing a large $M_\rho$ and a smaller $\lambda_1/(2\xi^2)$ it is very easy to make $m_X/f$ as small as possible, so the model itself does not provide a lower bound on $m_X/f$.  Since $f\gtrsim m_X$ in this model, the heavier top partner $m_{t_3}$ is expected to be at least 2 or 3 times $m_X$. On the other hand, higher $f$ requires more fine-tuning. For natural values of $f$, $m_X$ should not be very far above the current experimental bound.

Apart from the SM-like Higgs doublet, the other scalars in the model are heavy.  The masses of the heavy scalar doublet ($M_H$) and the CP-old singlet scalar are constrained to be larger than $K$, which needs to be large for small $f$ to retain an approximate custodial symmetry.  For $f\sim 1$~TeV, the constraint on the $T$ parameter requires $\tan{\beta} \lesssim 1.5$ (from Fig.~\ref{fig:mh1}), which gives $K^2 \gtrsim 1.2 \lambda_1 w^2  \gtrsim 1.7 m^2_X$ so that $ M_H\gtrsim 1.3 m_X$.  The CP-even (mostly) singlet scalar has a mass $\sim\sqrt{\lambda_1}f$, which is related to the mass of the heavier top partner $m_{t_3}\sim \xi f /\sqrt{2}$ by the standard NJL relation.  We have also assumed that the scalars in $\Sigma_{X,T,t}$ have masses much larger than $f$.  Therefore, the hypercharge $+7/6$ quarks $(X,T)$, being the lightest states in the model and carrying color, will be the first particles to be discovered if this model is realized in nature.  
Such hypercharge $+7/6$ quarks $(X,T)$ are a generic prediction of a composite Higgs model with a low chiral symmetry breaking scale and a custodial symmetry to avoid the $T$ parameter and $Zb\bar{b}$ coupling constraints. To unravel the underlying theory we would still need to find the other states and study their properties.
On the other hand, if the hypercharge $+7/6$ quarks $(X,T)$ are excluded up to a few TeV, in our model the chiral symmetry breaking scale would need to be at least as large, making the model as fine-tuned as the minimal model~\cite{Cheng:2013qwa}, then such an extension will be less motivated.  There have been many studies on the searches of charge $+5/3$ and $+2/3$ top partners \cite{Contino:2008hi, AguilarSaavedra:2009es, DeSimone:2012fs, Avetisyan:2013rca, Bhattacharya:2013iea, Barducci:2014ila}.  The estimated reach and exclusion regions on these quarks for the $14$ and $33$~TeV LHC can be found in the Snowmass 2013 report~\cite{Agashe:2013hma}, which is $\sim1.5(3)$~TeV for the $14(33)$~TeV LHC.  In any case, while it is possible to discover the $X$ and $T$ quarks at the $14$~TeV LHC, the $33$~TeV LHC or a future hadron collider is needed to probe the rest of the spectrum in our model.

The measurements of the couplings of the Higgs boson to SM particles provides an indirect way to probe or constrain models.  In our model, the Higgs boson has a small singlet component due to the mass mixings in Eq.~(\ref{eq:mmhhh}). As a result, the tree-level Higgs boson couplings to SM fields (except the top quark) are approximately reduced the by the factor $\cos{(v/f)} \approx 1-v^2/(2f^2)$, which is the fraction of the doublet component in the Higgs boson. For $f = 1$~TeV, the deviation from the SM couplings is $\sim 3\%$, which may be within the reach of a future $e^+e^-$ collider such as the ILC\cite{Peskin:2013xra}. The Higgs-top coupling can take a somewhat different value because the mixings of the top quark with vector-like quarks can induce additional contributions (of either sign). However, these contributions are very small within the viable parameter space because these top partners are quite heavy. The overall deviation of the Higgs-top coupling from its SM value is at most only a couple percents, similar to the other Higgs couplings.

\begin{figure}[!]
\centering
\includegraphics[height=8cm]{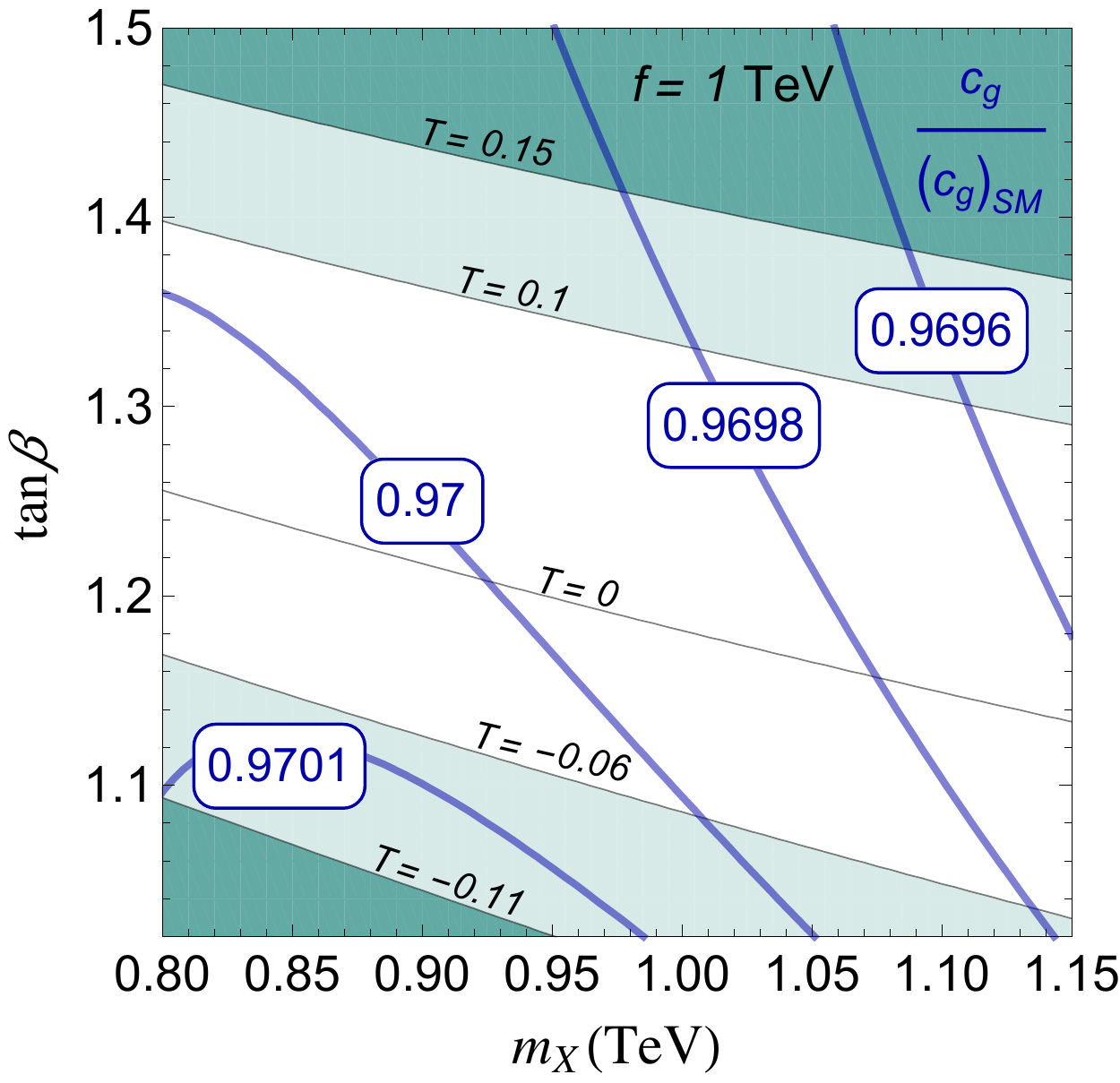}
\caption{ The ratio of the effective Higgs-gluon-gluon coupling to its SM value ($c_g/(c_g)_{\mbox{\scriptsize SM}}$) in the $(m_X,\tan{\beta})$ plane.  We fix $f=1$~TeV and choose the other parameters to be $\xi=3.6$, $\lambda_1/(2\xi^2)=0.7$, $\lambda_2/\lambda_1=0$, $M_\rho=3f$ and $M_{\Sigma_{X,T,t}}=10 f$.  This plot displays the same region of the parameter space as Fig.~\ref{fig:mh1}. }
\label{fig:cg}
\end{figure}

For the loop-induced Higgs couplings, there are additional contributions from the top partners. The Higgs production rate at the LHC would be modified if there is a sizable correction to the Higgs-gluon-gluon coupling.  In our model, the $X$ quark do not couple to the Higgs boson at tree level.  The three charge $+2/3$ quarks $t$, $T$ and $\chi$ mix and form mass eigenstates $t_1$, $t_2$ and $t_3$.  In the interesting region of the parameter space, the SM top-like  $t_1$ state has a similar coupling to the Higgs compared to the SM-value as discussed above.  The $t_2$ and $t_3$ states receive most of their masses from the electroweak-preserving vector-like mass terms. Their couplings to the Higgs boson come from mixings and are highly suppressed.  The effective Higgs-gluon-gluon coupling ($c_g$) can be calculated by integrating out the heavy fermions $t_1$, $t_2$ and $t_3$ in the loops. The result for $f=1$~TeV is shown in Fig.~\ref{fig:cg} (with the same parameters as in Fig.~\ref{fig:mh1}). For these parameters $c_g/(c_g)_{\mbox{\scriptsize SM}} \approx 0.97$, very close to corrections of other Higgs couplings.  The $3\%$ deviation is much smaller than the current LHC bound, but could be probed by a future $e^+e^-$ collider.

%


\section{Conclusions}
\label{sec:con}

Top seesaw models are a natural framework to incorporate the composite Higgs as the pNGB of the broken chiral symmetry which relates the SM top, bottom quarks and new vector-like quarks. The Higgs boson mass is strongly correlated with the top quarks mass because both come from the same explicit chiral symmetry breaking effect. Consequently, the 126 GeV Higgs is easily accommodated within natural range of model parameters.  This type of models also have a decoupling limit where the standard model is recovered in the limit of large chiral symmetry breaking scale, albeit with a price of fine tuning. A natural model should have the chiral symmetry breaking scale not far above the weak scale. However, it will potentially give large corrections to the SM observables, and hence subjects to strong experimental constraints. In the minimal top seesaw model where only one vector-like singlet quark $\chi$ is added, the strongest constraint comes from the weak-isospin violation $T$ parameter. The $U(3)_L$ chiral symmetry among $(t_L, b_L, \chi_L)$ does not contain a custodial symmetry to protect the weak-isospin. As a result, a large contribution to $T$ is generated from the fermion loops if the chiral symmetry breaking scale is close to the weak scale. The experimental constraint pushes the chiral symmetry breaking scale above 3.5 TeV, implying a strong fine-tuning in that model. The new states associated with the model are also beyond the reach of the LHC.

In this paper we studied an extension of the top seesaw model that contains a custodial symmetry to evade the strong constraint from the $T$ parameter. Although a simple extension to include the bottom seesaw can protect the weak-isospin, it suffers from the constraint on the $Z b \bar{b}$ coupling due to the mixing of the $b_L$ with a EW singlet quark. To avoid both problems, in addition to the usual singlet vector-like top partner, we need to introduce vector-like EW doublet quarks, $(X, T)$, with $(X_L, T_L, t_L, b_L)$ forming a $(2,\, 2)$ representation of the $O(4) \simeq SU(2)_L\times SU(2)_R$ symmetry. With these additional quarks, the strong dynamics which form composite scalars may have an enlarged symmetry. We showed that if the theory preserves an approximate $O(5)$ symmetry below the compositeness scale, an SM-like Higgs doublet arises naturally as the pNGB of the $O(5)\to O(4)$ symmetry breaking, while the remaining $O(4)$ contains an approximate custodial symmetry to protect the weak-isospin. There is also no large shift of $b_L$ coupling to $Z$ since there is no mixing with a EW singlet quark. As a result, the chiral symmetry breaking scale $f$ can be significantly lowered while satisfying all EW precision constraints. The lower bound on $f$ in this model comes from the search of the electric charge 5/3 $X$ quark at the LHC. To produce the Higgs boson mass at 126 GeV, we found that $f$ needs to be somewhat larger than the $X$ quark mass. The current LHC bound on $X$ quark mass of 800 GeV renders a lower bound on $f$ of the order of 1 TeV. The tuning, measured by $v^2/f^2$, can be improved to $\sim 5\%$, compared to $\lesssim 0.5\%$ in the minimal model.

Naturalness does not come without a price.  To reduce fine-tuning and to avoid the experimental constraints, we are forced to introduce the $X$ and $T$ quarks and the corresponding composite scalars, making the structure of the theory much more complicated.  As a matter of fact, the minimal model in Ref.~\cite{Cheng:2013qwa} and the extended model studied in this paper are another example of the so-called ``crossroads" situation, and one has to choose between fine-tuning and complexity.  Ultimately, both models need to be tested by experiments.  The search for the $X$ and $T$ quarks at the 14~TeV (and possibly the 33~TeV) LHC can provide important clues in discriminating the two scenarios.  However, to fully probe either model, one needs to go beyond the LHC.  There has been discussion of a future 100 TeV Hadron Collider that could be built either at CERN~\cite{cern100TeV} or in China~\cite{china100TeV}.  Such a collider, if realized, will further probe the origin of the EWSB and tell us which road our Mother Nature takes.


\subsection*{Acknowledgments}
We would like to thank Bogdan Dobrescu and Ennio Salvioni for discussion.    This work is supported by the Department of Energy (DOE) under contract no.\ DE-FG02-91ER40674. 


\appendix



\section{T parameter from fermion loops}
\label{app:T}

In Section \ref{sec:fermion} we argued that the leading contribution to the $T$ parameter is captured by the fermion loops.  In this appendix, we provide an expression for the $T$ parameter calculated from fermion loops.  In terms of $SU(2)_W$ eigenstates, the contribution comes from the fermions $(t_L,b_L)$, $(X_L, T_L)$ and $(X_R, T_R)$ [since  $(X_R, T_R)$ is also a $SU(2)_W$ doublet].  The charge $+2/3$ fermions, $t$, $T$ and $\chi$ form a $3\times 3$ mass matrix, as shown in Eq.~(\ref{eq:fermass}).  We denote the three mass eigenstates as $t_1$, $t_2$ and $t_3$, ordered by $m_{t_1} \leq m_{t_2} \leq m_{t_3}$, and denote the left-handed and right-handed rotation matrices as
\begin{equation}
\bpm t_{1L} \\ t_{2L} \\ t_{3L} \epm = \bpm L_{11} & L_{12} & L_{13} \\  L_{21} & L_{22} & L_{23} \\  L_{31} & L_{32} & L_{33} \epm \bpm t_L \\ T_L \\ \chi_L \epm ~, \hspace{0.5cm}
\bpm t_{1R} \\ t_{2R} \\ t_{3R} \epm = \bpm R_{11} & R_{12} & R_{13} \\  R_{21} & R_{22} & R_{23} \\  R_{31} & R_{32} & R_{33} \epm \bpm t_R \\ T_R \\ \chi_R \epm ~.
\end{equation}

The contribution to the $T$ parameter from fermion loops is 
\begin{equation}
T=\frac{3}{16\pi^2 \alpha v^2} \left[A+B-C   \right]~,
\label{eq:T}
\end{equation}
where
\begin{align}
A =&~ 2m^2_X +m^2_b +\left[ (L^2_{11}-L^2_{12})^2+R^4_{12}  \right] m^2_{t_1} \nonumber\\
& +\left[ (L^2_{21}-L^2_{22})^2+R^4_{22}  \right] m^2_{t_2} +\left[ (L^2_{31}-L^2_{32})^2+R^4_{32}  \right] m^2_{t_3}  \nonumber\\
& -L^2_{11} f(m_{t_1},m_b) -L^2_{21} f(m_{t_2},m_b) -L^2_{31} f(m_{t_3},m_b) \nonumber\\
& -(L^2_{12}+R^2_{12}) f(m_X, m_{t_1}) -(L^2_{22}+R^2_{22}) f(m_X, m_{t_2}) -(L^2_{32}+R^2_{32}) f(m_X, m_{t_3}) \nonumber\\
& +\left[ (L_{11}L_{21}-L_{12}L_{22})^2+R^2_{12}R^2_{22} \right] f(m_{t_1},m_{t_2}) \nonumber\\
& +\left[ (L_{11}L_{31}-L_{12}L_{32})^2+R^2_{12}R^2_{32} \right] f(m_{t_1},m_{t_3}) \nonumber\\
& +\left[ (L_{21}L_{31}-L_{22}L_{32})^2+R^2_{22}R^2_{32} \right] f(m_{t_2},m_{t_3})  ~,
\end{align}
\begin{align}
B =&~ 2L_{12}R_{12} \,g(m_X,m_{t_1}) + 2L_{22} R_{22} \,g(m_X, m_{t_2}) + 2L_{32}R_{32}\,g(m_X,m_{t_3}) \nonumber\\
& + (L^2_{11}-L^2_{12})R^2_{12} \,g(m_{t_1},m_{t_1}) + (L^2_{21}-L^2_{22})R^2_{22} \,g(m_{t_2},m_{t_2}) \nonumber\\
& + (L^2_{31}-L^2_{32})R^2_{32} \,g(m_{t_3},m_{t_3}) - g(m_X,m_X) \nonumber\\
& + 2(L_{11}L_{21}-L_{12}L_{22})R_{12}R_{22} \,g(m_{t_1},m_{t_2}) \nonumber\\
& + 2(L_{11}L_{31}-L_{12}L_{32})R_{12}R_{32} \,g(m_{t_1},m_{t_3}) \nonumber\\
& + 2(L_{21}L_{31}-L_{22}L_{32})R_{22}R_{32} \,g(m_{t_2},m_{t_3}) ~,
\end{align}
and
\begin{equation}
C = m^2_{\rm top} + m^2_b + f(m_{\rm top}, m_b)
\end{equation}
is the contribution from the Standard Model $(t_L,b_L)$ that needs to be subtracted.  In our model, the top quark is always the lightest eigenstate of the top mass matrix, i.e.\ $m_{\rm top}=m_{t_1}$.  The functions $f$ and $g$ in the above expressions are given by 
\begin{align}
f(a,b) &= \frac{2a^2 b^2}{a^2-b^2}\log{(\frac{a^2}{b^2})} ~, \\
g(a,b) &= 4ab(-1+\frac{a^2\log{a^2}-b^2\log{b^2}}{a^2-b^2}) ~,
\end{align}
while in the limit that $a=b$,
\begin{equation}
f(a,a)=2a^2~, \hspace{1cm} g(a,a)=4a^2 \log{a^2}~.
\end{equation}
Eq.~(\ref{eq:T}) is used in the numerical studies presented in Section~\ref{sec:plots}.   


\section{Estimation of coupling ratios $\lambda_1/(2\xi^2)$ and $\lambda_2/\lambda_1$}
\label{app:rge}

The predictions of our model depend on the values of the Yukawa coupling $\xi$ in Eq.~(\ref{eq:Yukawa}) and quartic couplings $\lambda_1, \, \lambda_2$ in Eq.~(\ref{eq:Vl2}).  It was shown in the previous paper\cite{Cheng:2013qwa}  that the ratios of couplings, $\lambda_1/(2\xi^2)$ and $\lambda_2/\lambda_1$, are better estimated than their individual values. At the same time the predictions of the model, such as the mass of the Higgs boson, also have stronger dependences on the ratio of the couplings.  This is also true for the model studied in this paper.  With the addition of the $(X,T)$ quarks, the estimated coupling ratios are slightly modified from the minimal model~\cite{Cheng:2013qwa}, while the derivations remain the same.  Here we provide a short summary of the results and refer the reader to the appendix of \cite{Cheng:2013qwa} for more details of this study.

In the fermion bubble approximation, the ratio $\lambda_1/(2\xi^2)$ is predicted to be $1$, while $\lambda_2$ is zero since it is not generated by the fermion loops.  These results are modified once the gauge loop corrections and the back reaction of the scalar self-interactions are included, for example, by using the full one loop RG equations~\cite{Bardeen:1989ds}.  If the chiral symmetry breaking scale $f$ is not much smaller than the compositeness scale $\Lambda$, as in the case that we are interested, one cannot trust the RG analysis because the couplings are strong and the logarithms are only ${\cal O}(1)$.  Nevertheless, it may provide us some ideas of the possible range of the coupling ratios $\lambda_1/ (2 \xi^2)$ and $\lambda_2/\lambda_1$.

The coupled RG equations of the couplings $\xi$, $\lambda_1$, $\lambda_2$, and QCD strong coupling $g_3$ for an $U(N_L)_L \times U(N_R)_R$ theory are given by
\begin{eqnarray}
16\pi^2 \frac{d g_3}{d t} &=& - \left(11-\frac{2}{3}N_f \right) g^3_3\, ,   \nonumber\\
16\pi^2 \frac{d\xi}{dt} &=& \left(\frac{N_L+N_R}{2}+ N_c\right) \xi^3- 3\, \frac{N_c^2 -1}{N_c} g^2_3 \, \xi\, ,    \nonumber\\
16\pi^2 \frac{d \lambda_1}{dt} &=& 2(N_L+N_R) \lambda^2_1 + 4\lambda_1\lambda_2 + 4 N_c (\xi^2\lambda_1 - \xi^4)\, ,   \nonumber\\
16\pi^2 \frac{d \lambda_2}{dt} &=& 4\lambda^2_1 + 4(N_L+N_R) \lambda_1\lambda_2 + 2 N_L N_R \lambda^2_2 + 4 N_c\xi^2\lambda_2\, , \label{eq:rg1}
\end{eqnarray}
\begin{figure}[t]
\centering
\includegraphics[width=10cm]{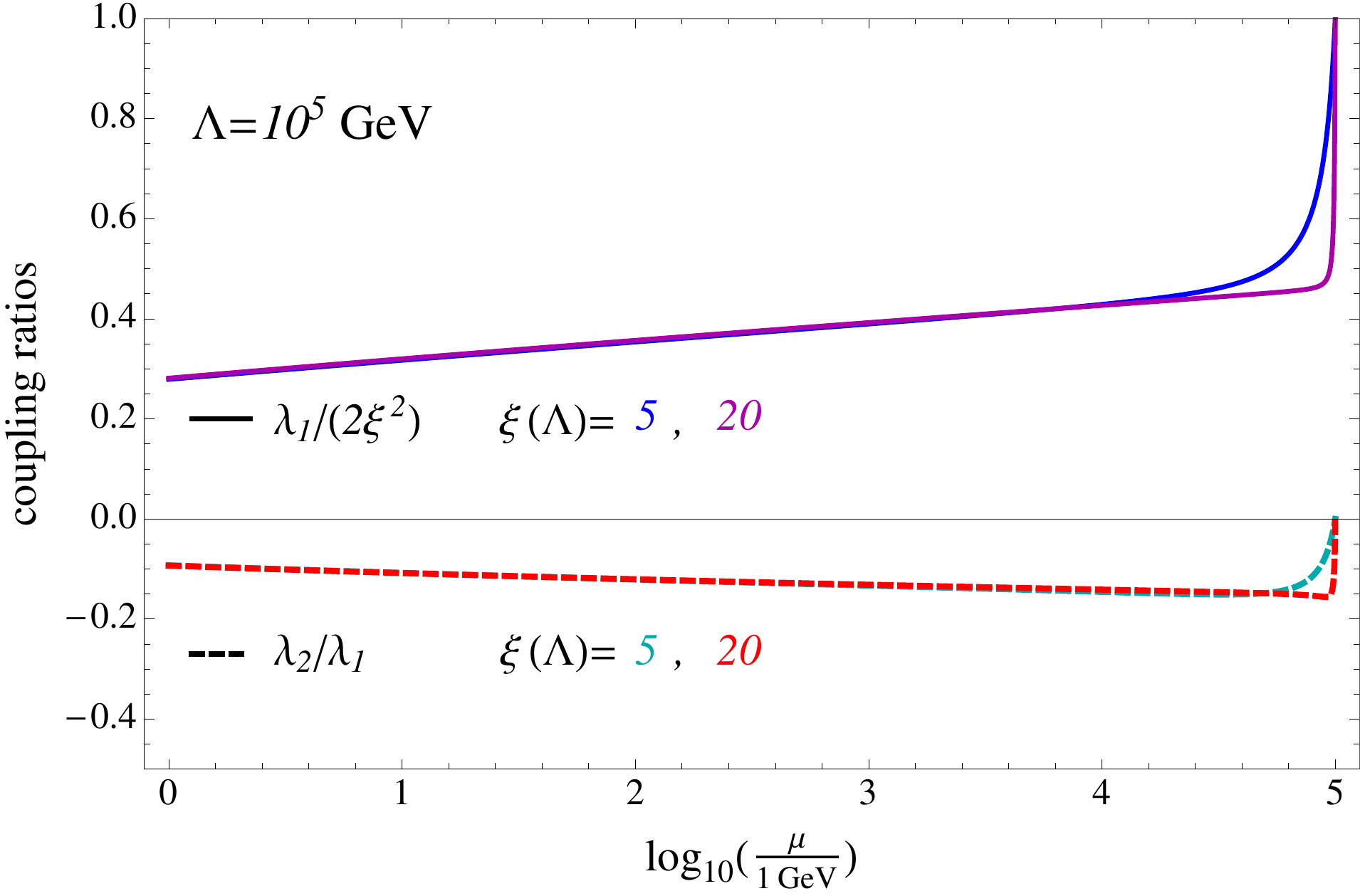}
\caption{One-loop RG evolutions of the coupling ratios $\lambda_1/(2\xi^2)$ and $\lambda_2/\lambda_1$ for initial values $\lambda_1/(2\xi^2)=1$, $\lambda_2/\lambda_1=0$ and $\xi=5$ or 20. The horizontal axis is the logarithm of the energy scale.  The following values are used: $N_L=5,\, N_R=4$, $N_c=3$ and $N_f = 9$.  }
\label{fig:rg}
\end{figure}
where we have ignored the EW couplings $g_1$, $g_2$, and the light fermion Yukawa couplings. $N_f$ is the number of quark flavors.  We solve these equations numerically for our model which has $N_L=5,\, N_R=4$, $N_c=3$ and $N_f = 9$. We set the initial conditions $\lambda_1 = 2 \xi^2, \, \lambda_2=0$ and choose several different initial values for $\xi$. 

The results are shown in Fig.~\ref{fig:rg}.  The ratios of couplings are quickly driven to some approximate fixed point values, though we should not trust the exact evolution near $\Lambda$ due to potentially large higher loop contributions.

If the chiral symmetry breaking scale is not far below the compositeness scale, we can not trust the 1-loop RG results. However, if we assume a smooth evolution, the ratios of couplings are expected to lie in between their initial values and the quasi-infrared fixed point values:
\begin{equation}
0.35 \lesssim \frac{\lambda_1}{2 \xi^2} \lesssim 1, \quad -0.15 \lesssim \frac{\lambda_2}{\lambda_1} \lesssim 0 \, .
\label{eq:ratio_range}
\end{equation}
We adopt these ranges in Section~\ref{sec:let}~and~\ref{sec:plots}.

\providecommand{\href}[2]{#2}\begingroup\raggedright\endgroup

\end{document}